\documentclass[aps,prev,onecolumn,preprintnumbers,floatfix,nofootinbib]{revtex4-1}
\pdfoutput=1
\usepackage[utf8x]{inputenc}
\usepackage{ucs}
\usepackage[english]{babel}
\usepackage[T1]{fontenc}
\usepackage{amsmath}
\usepackage{amsfonts}
\usepackage{amssymb}
\usepackage[top=2.95 cm, bottom=2.95 cm, left=2.95 cm, right= 2.95 cm ]{geometry}
\usepackage{wrapfig}
\usepackage{graphicx,bbm,mathrsfs}  
\usepackage{subfig}
\usepackage{epsfig}
\usepackage{rotating}
\usepackage{dcolumn}
\usepackage{bm}
\usepackage{float}
\usepackage{hyperref}
\usepackage{pstricks}
\usepackage{color}
\usepackage{verbatim}

\newcommand{\Ref}[1]{Ref.~\cite{#1}}

\newcommand{\beq}{\begin{equation}} 
\newcommand{\eeq}{\end{equation}}

\begin{document}

\preprint{LPT-Orsay-16-77}

\vspace*{1mm}

\title{Dark Matter Phenomenology of SM and Enlarged Higgs Sectors Extended with Vector Like Leptons}

\author{Andrei Angelescu$^{a}$}
\email{andrei.angelescu@th.u-psud.fr}
\author{Giorgio Arcadi$^{a,b}$}
\email{giorgio.arcadi@mpi-hd.mpg.de}

\vspace{0.1cm}
\affiliation{
${}^a$ 
Laboratoire de Physique Th\'eorique, CNRS,
Universit\'e Paris-Saclay, 91405 Orsay, France
}
\affiliation{
${}^b$ 
Max Planck Instit\"{u}t f\"{u}r Kernphysik, Saupfercheckweg 1, 69117 Heidelberg, Germany
}

\begin{abstract} 

\noindent
We will investigate the scenario in which the Standard Model (SM) Higgs sector and its 2-doublet extension (called the Two Higgs Doublet Model or 2HDM) are the ``portal'' for the interactions between the Standard Model and a fermionic Dark Matter (DM) candidate. The latter is the lightest stable neutral particle of a family of vector-like leptons (VLLs). We will provide an extensive overview of this scenario combining the constraints purely coming from DM phenomenology with more general constraints like Electro-weak Precision Test (EWPT) as well as with collider searches. In the case that the new fermionic sector interacts with the SM Higgs sector, constraints from DM phenomenology force the new states to lie above the TeV scale. This requirement is relaxed in the case of 2HDM. Nevertheless, strong constraints coming from Electroweak Precision Tests (EWPT) and the Renormalization Group Equations (RGEs) limit the impact of VLFs on collider phenomenology.

\end{abstract}

\maketitle

\tableofcontents

%%%%%%%%%%%%%%%%%%%%%%%%%%%%%%%%%%%%%%%%%%%%%%%%%%%%%%%%%%%%%%%%%%%%%%%%%%
\section{Introduction}
%%%%%%%%%%%%%%%%%%%%%%%%%%%%%%%%%%%%%%%%%%%%%%%%%%%%%%%%%%%%%%%%%%%%%%%%%%

\noindent
Weakly Interacting Massive Particles (WIMPs) represent probably the most popular class of Dark Matter (DM) candidates. Among the features which make this kind of candidates so attractive, it is for sure worth mentioning the production mechanism. WIMP DM were indeed part of the primordial thermal bath at Early stages of the history of the Universe and decoupled (freeze-out) at later stages, when the temperature was below their mass (i.e. non-relativistic decoupling), since the interactions with the SM particles were not efficient anymore with respect to the Hubble expansion rates. Under the assumption of standard cosmological origin, the comoving abundance of the DM, which is determined by the one at freeze-out, is set by a single particle physics input, namely the thermally averaged pair annihilation cross-section. The experimentally favored value of DM abundance, expressed by the quantity $\Omega h^2 \approx 0.12$~\cite{Ade:2015xua} corresponds to a thermally averaged cross-section $\langle \sigma v \rangle \sim 10^{-26}{\mbox{cm}}^3 {\mbox{s}}^{-1}$. Interactions of this size are potentially accessible to a broad variety of search strategies, ranging from Direct/Indirect Detection to production at colliders, making the WIMP paradigm highly testable.

\noindent
From the point of view of model building, WIMP frameworks feature interactions between pairs of Dark Matter particles (in order to guarantee the cosmological stability of the DM, operators with a single DM field are in general forbidden, e.g. through a symmetry) and pair of SM states, induced by suitable mediator fields. The simplest option, in this sense, is probably represented by s-channel electrically neutral mediators, dubbed ``portals'', which can couple the DM with SM fermions (see e.g.~\cite{Abdallah:2014hon,Kahlhoefer:2015bea,Arina:2016cqj}), although couplings with the SM gauge bosons might also be feasible~\cite{Chu:2012qy,Mambrini:2015wyu,Backovic:2015fnp,D'Eramo:2016mgv}. The DM relic density is thus determined via s-channel exchange of the mediator states. By simple crossing symmetry arguments these processes can be, for example, related to the rate of DM Direct Detection, induced by the t-channel interaction between the DM and the SM quarks, and to the ones of DM pair production at colliders, which can be probed mostly through mono-jet events~\cite{Goodman:2010ku,Fox:2011pm,Aad:2015zva,Khachatryan:2014rra}.

\noindent
Interestingly, the SM features two potential s-channel mediators, namely the $Z$ and the Higgs bosons. One possible result are ``Z-portal'' DM~\cite{Arcadi:2014lta} scenarios. However, they are rather contrived, since, because of gauge invariance, interaction between a SM singlet DM and the $Z$ can arise only at the non-renormalizable level~\cite{Cotta:2012nj,deSimone:2014pda}. ``Higgs portal'' models are instead very popular, although rather constrained~\cite{Lebedev:2011iq,Djouadi:2011aa,Djouadi:2012zc,LopezHonorez:2012kv,Beniwal:2015sdl}, since a DM spin-0 (1), even if it is a singlet with respect to the SM gauge group, can interact with the SM Higgs doublet $H$ via four-field operators connecting the bilinear $H H^\dagger$ with a DM pair and giving rise, after electroweak (EW) symmetry breaking, to an effective vertex between a DM pair and the physical Higgs field $h$. 

\noindent
The fermionic ``Higgs portal'' is instead a dimension-5 operator. Furthermore this is strongly constrained, also with respect to the scalar and vector DM cases, because of the strong direct detection rates accompanied by a velocity suppressed annihilation cross-section~\cite{Djouadi:2012zc,LopezHonorez:2012kv}.

\noindent
In order to couple, at the renormalizable level, the $Z$ and/or the Higgs bosons with a fermionic DM, the latter should feature a (small) hyper- and/or $SU(2)$ charge. This could be realized through the mixing of a pure SM singlet and extra states with non-trivial quantum numbers under $SU(2) \times U(1)$.

\noindent
A concrete realization consists in the introduction of a family of vector-like leptons (VLLs) (see e.g.~\cite{Cohen:2011ec,Calibbi:2015nha,Yaguna:2015mva,Berlin:2015wwa} for alternatives). This gives a set of vector like fermions (VLFs) with the same quantum numbers of the SM leptons and of the right-handed neutrinos. In absence of mixing with SM leptons, the lightest new fermionic state, if electrically neutral, constitutes a DM candidate. We also notice that this kind of scenario presents a richer phenomenology with respect to the simple Higgs portal. In addition, the DM achieves, through the aforementioned mixing, hypercharge and weak isospin such that it features non-zero interactions with the $W$ and $Z$ bosons. 

\noindent
SM extensions with one family of VLLs are constrained, from the point of view of DM phenomenology, in a totally analogous way as Higgs and Z-portal models. This is mainly because the Higgs and the Z-boson are responsible of Spin Indepedent (SI) interactions between the DM and nucleons, which result in increasing tension with experimental bounds from Direct Detection experiments. As will be shown in the following it is possible to comply with these limits and achieve, at the same time, the correct relic density only for DM masses above the TeV scale.

\noindent
A more interesting scenario is obtained if the Higgs sector is extended by a second doublet (2HDM). In this case it is possible to comply with Direct Detection limits at lower values of the DM mass. This result is achieved in two ways. In the low, i.e. lighter than the Higgs scalars, DM mass regime, its pair annihilation cross-section is enhanced by s-channel resonances. For higher values of the DM masses the enhancement is instead provided by new annihilation channels with the extra Higgs bosons as final states.

\noindent
2HDM+VLFs models have attracted great attention in the recent times since they allowed for the interpretation of the 750 GeV diphoton excess~\cite{Angelescu:2015uiz,Djouadi:2016eyy,Benbrik:2015fyz,Murphy:2015kag,Wang:2015omi,Tang:2015eko,Dev:2015vjd,Palti:2016kew,Dutta:2016jqn,DiChiara:2016dez,Kawamura:2016idj,Gopalakrishna:2016tku,Arhrib:2016rlj}, announced by the LHC collaboration in December 2015~\cite{ATLASdiph,CMS:2015dxe,CMS:2016owr,Moriondatlas}, but not confirmed by the 2016 data~\cite{ATLAS1,CMS1}.

\noindent
In this work, we will review this kind of scenarios focusing on the solution to the DM problem. As already mentioned, we will consider a specific set of VLFs, consisting into a family of vector-like leptons. We will analyze the most relevant aspects of the DM phenomenology, determining the regions providing the correct relic density, according to the WIMP paradigm, and verifying whether these regions are compatible with the constraints from DM experimental searches, especially DM Direct Detection.

\noindent
The parameter of the new fermion sector are, however, constrained not only by DM phenomenology. The size of their couplings to the 125 GeV Higgs is instead constrained by Electro-Weak Precision Tests (EWPT). A further strong upper bound on these couplings, as well as the ones with the other Higgs states, comes from the RG running of the the gauge and the quartic couplings of the scalar potential. In particular, the latter get strong negative contributions proportional to the fourth power of the yukawa couplings of the VLLs, such that the scalar potential might be destabilized even at collider energy scales, unless new degrees of freedom are added. Finally, VLLs can give provide ``indirect'' collider signals, through possible modification of diphoton production rates, as well as possible signals associated to their direct production.

\noindent
This set of constraints, together with the ones from conventional searches of extra Higgs scalars in collider and low energy processes are complementary to the ones coming from DM phenomenology. Their combination can be used to determine the range of viable DM masses (an in turn provide information on the rest of the VLF mass spectrum) for several 2HDM realizations.

\noindent
The paper is organized as follows. We will firstly introduce, at the beginning of section II, the ``family'' of vector-like fermions. The remainder of the section will be dedicated to a brief overview of the SM+VLLs scenario. Firstly, we will briefly illustrate the general constraints coming from the modification of the Higgs signal strengths and the Electroweak Precision Tests (EWPT), and afterwards focus on the DM phenomenology. Along similar lines, an analysis for the 2HDM will then be performed in section III. After a short review of the general aspects of 2HDMs, we will perform a more detailed analysis of the constraints from EWPT and Higgs signal strengths and add to them the RGE constraints. After the analysis of the DM phenomenology, we will briefly discuss the limits/prospects, for our scenario, of collider searches. Finally, we will summarize our results in section III.G and conclude in section IV.

%%%%%%%%%%%%%%%%%%%%%%%%%%%%%%%%%%%%%%%%%%%%%%%%%%%%%%%%%%%%%%%%%%%%%%%%%%
\section{Vector-Like Extensions of the Standard Model}
\label{SM}
%%%%%%%%%%%%%%%%%%%%%%%%%%%%%%%%%%%%%%%%%%%%%%%%%%%%%%%%%%%%%%%%%%%%%%%%%%
\noindent
In this section we will review how introducing vector-like leptons affects the SM Higgs sector. As already pointed out, the impact is mostly twofold. First of all, they generate additional loop contributions to the loop-induced couplings of the Higgs boson to two photons, giving rise to deviations of the corresponding signal strength with respect to the SM prediction. In addition, the presence of vector-like leptons is typically associated with sensitive departures from experimental limits for the EW precision observables. In order to have viable values of the Higgs signal strengths and precision observables, one should impose definite relations for the Yukawa couplings and masses of the new VLLs. The same relations will hold, up to slight modifications, also in the 2HDM case.

%%%%%%%%%%%%%%%%%%%%%%%%%%%%%%%%%%%%%%%%%%%%%%%%%%%%%%%%%%%%%%%%%%%%%%%%%%
\subsection{The Vector-Like ``Family''}
\label{SM-family}
%%%%%%%%%%%%%%%%%%%%%%%%%%%%%%%%%%%%%%%%%%%%%%%%%%%%%%%%%%%%%%%%%%%%%%%%%%

\noindent
In this work we will assume that the SM and, afterwards, the 2HDM Higgs sectors can be extended by ``families'' of vector like fermions (VLFs). By family we understand a set of two $SU(2)_L$ singlets and two $SU(2)_L$ doublets, belonging to a $SU(3)_c$ representation $R_c$, and with their hypercharge determined by a single parameter, $Y$. For the moment, we will keep the discussion general and later on specialize on possible DM candidates. The new fields can be schematically labeled as:
\begin{equation}
{\cal D}_{L,R}  \sim (R_c, 2, Y-1/2)~, 
\quad
U^\prime_{L,R} \sim  (R_c, 1, Y)~,\quad
 D^\prime_{L,R} \sim  (R_c, 1, Y-1)~,
 \label{general-VL-family}
\end{equation}
so that the couplings to the SM Higgs doublet, $H={\left(0 \; \frac{v+h}{\sqrt{2}}\right)}^{T}$, are parametrized by the following Lagrangian:
\begin{eqnarray}
-{\cal L}_{\rm VLF} &=& y^{U_R} \overline{{\cal D}_L} \tilde{H} U^\prime_R + y^{U_L} \overline{U^\prime_L} \tilde{H}^\dagger {\cal D}_R +
y^{D_R} \overline{{\cal D}_L} H D^\prime_R + y^{D_L} \overline{D^\prime_L} H^\dagger {\cal D}_R 
\nonumber \\
&&
+M_{UD} \overline{{\cal D}_L} {\cal D}_R 
+ M_U \overline{U^\prime_L} U^\prime_R +M_D \overline{D^\prime_L} D^\prime_R + {\rm h.c.}~,
\end{eqnarray}
where we have considered the following decomposition for the $SU(2)$ doublets:
${\cal D}_{L,R}\equiv
\begin{pmatrix}
U & D
\end{pmatrix}^T_{L,R}$.

\noindent
For simplicity we will assume that all the couplings are real and that the mixing between the VLFs and the SM fermions is negligible.

\noindent
After electroweak symmetry breaking (EWSB), there is a mixing in the ``up'' ($U', U$) and ``down'' ($D', D$) sectors.
The ``up'' VL fermions have charge $Q_U=Y$, while the ``down'' fermions have charge $Q_D=(Y-1)$. 
The mass matrices in the two sectors are
\begin{equation}
{\cal M }_U= 
\begin{pmatrix}
M_U & y^{U_L} v/ \sqrt{2}
\\
y^{U_R} v/ \sqrt{2} & M_{UD}
\end{pmatrix}
~,
\qquad\qquad
{\cal M}_D=
\begin{pmatrix}
M_D &  y^{D_L} v/ \sqrt{2}
\\
y^{D_R} v/ \sqrt{2} & M_{UD}
\end{pmatrix}~,
\label{SM-mass-matrices}
\end{equation}
with $v=246$~GeV, and they are bi-diagonalized as follows
\begin{equation}
U_L^{F} \cdot {\cal M}_{F} \cdot \left(U_R^{F} \right)^\dagger =
\begin{pmatrix}
m_{F_1} & 0\\
0 & m_{F_2}
\end{pmatrix}
~,
\quad
U_L^{F} =
\begin{pmatrix}
c_L^{F} & s_L^{F} \\
-s_L^{F} & c_L^{F}
\end{pmatrix}
~,
\quad
U_R^{F}=
\begin{pmatrix}
c_R^{F} & s_R^{F} \\
-s_R^{F} & c_R^{F}
\end{pmatrix}~,
\end{equation}
where the sub/superscripts $F=U,D$ distinguish between the two sectors and $c_{L/R}^F = \cos \theta_{L/R}^F$, $s_{L/R}^F = \sin \theta_{L/R}^F$. Throughout this work we will denote the lighter  mass eigenstate as $F_1$. The limit where one of the singlets is decoupled, e.g. when $y_{U_R}=y_{U_L}=0$ and $M_D\rightarrow \infty$, has already been studied in detail in \Ref{Bizot:2015zaa}.
As we will see below the mixing structure in Eq.~\ref{SM-mass-matrices} is strongly constrained by the electroweak precision tests (EWPT) and by the Higgs couplings measurements.

%%%%%%%%%%%%%%%%%%%%%%%%%%%%%%%%%%%%%%%%%%%%%%%%%%%%%%%%%%%%%%%%%%%%%%%%%%
\subsection{Electroweak Precision Tests}
\label{SM-EWPT}
%%%%%%%%%%%%%%%%%%%%%%%%%%%%%%%%%%%%%%%%%%%%%%%%%%%%%%%%%%%%%%%%%%%%%%%%%%

\noindent
Extending the SM with vector-like fermions leads, in general, to the deviation of the Electroweak precision observables $S$ and $T$ from their respective experimental limits. Assuming negligible mixing between the SM and the vector-like fermions, the limits on $S$ and $T$ can be directly translated into limits on the Yukawa couplings and masses of the new fermions; in the limit in which the former go to zero, no constraints from EWPT apply. 

\begin{figure}[h]
\begin{center}
\includegraphics[keepaspectratio=true,width=0.5\textwidth]{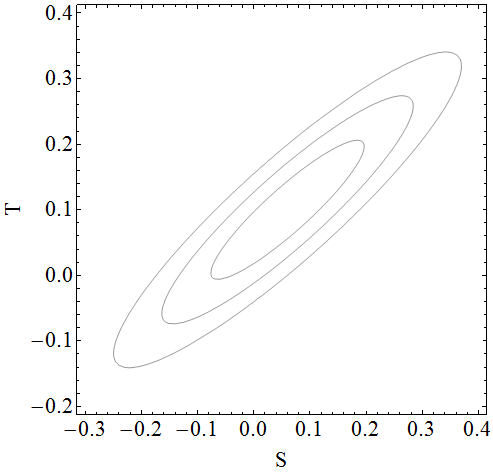}
\caption{
{\small Allowed values of S and T~\cite{Baak:2011ze} at, from the innermost to the outermost ellipse, 68\%, 95.5\% and 99.7\% confidence level (CL).}  
}
\label{Fig:st_ellipse}
\end{center}
\end{figure}

\noindent
Sizable values of the Yukawa couplings of the VLFs can nevertheless be obtained while still complying with the limits on the $T$ parameters by relying (at least approximately) on a custodial limit:
\begin{equation}
M_D=M_U~,
\qquad
y^{U_L}=y^{D_L}~,
\qquad
y^{U_R}=y^{D_R}~,
\end{equation}
which is equivalent to imposing equal mass matrices in the isospin-up and isospin-down sectors. Clearly, the custodial limit can be achieved only by considering ``full families'' of VLFs, i.e. a corresponding SU(2) singlet for each of the components of the doublet, as done in this work.
\noindent
On the contrary, there is no symmetry protecting the $S$ parameter, which means that, in some cases, it will impose more relevant constraints than the $T$ parameter. The constraints on $S$ can be nevertheless partially relaxed by taking advantage of the correlation among the $S$ and $T$ parameters, illustrated in fig.~\ref{Fig:st_ellipse}, by allowing for a small deviation from the custodial limit, i.e. $T \gtrsim 0$.

%%%%%%%%%%%%%%%%%%%%%%%%%%%%%%%%%%%%%%%%%%%%%%%%%%%%%%%%%%%%%%%%%%%%%%%%%%
\subsection{Higgs Couplings}
\label{SM-Higgs}
%%%%%%%%%%%%%%%%%%%%%%%%%%%%%%%%%%%%%%%%%%%%%%%%%%%%%%%%%%%%%%%%%%%%%%%%%%

\noindent
We now turn to the second constraint coming from the Higgs couplings measurements. In the presence of vector-like fermions, its couplings to gauge bosons receive additional contributions, originating from triangle loops in which the new fermions are exchanged. No new decay channels into VLFs are instead present since, because of constraints from direct searches at colliders, the VLFs should be heavier than the SM Higgs.

\noindent
The SM Higgs loop-induced partial decay widths into massless gauge bosons, $\Gamma_{h \rm{VV}},\, \mathrm{V}=g,\gamma$, can be schematically expressed as $\Gamma_{h\rm{VV}} \propto |\mathcal{A}_{\rm SM}^{h\rm{VV}}+\mathcal{A}_{\rm VLF}^{h\rm{VV}}|^2$, where $\mathcal{A}_{\rm SM}^{h\rm{VV}}$ and $\mathcal{A}_{\rm VLF}^{h\rm{VV}}$ represent the amplitudes associated, respectively, to the SM and VLF contributions. Throughout this work we will only consider the case of a family of color-neutral VLFs ($R_c=1$); as a consequence the new physics sector influences mostly $\Gamma_{h\gamma \gamma}$ and therefore the $h\to\gamma\gamma$ signal strength, $\mu_{\gamma\gamma}$.\footnote{Note that $\mu_{hZ\gamma}$ is also affected by the VLFs, but the uncertainties on this signal strength are too large to constrain the extended fermionic sector~\cite{Aad:2015gba,Khachatryan:2014jba}.} The corresponding amplitude is given by:
\begin{equation}
{\cal A}^{h\gamma \gamma}_{\rm VLF}=
\sum_{\substack{F=U,D \\ i=1,2}} Q_F^2 \frac{v (\mathcal{C}_F)_{ii}}{m_{F_i}}A_{1/2}^h (\tau^h_{F_i}),  
\label{SM-App}
\end{equation}
where $\tau^h_{F_i}=\frac{m_h^2}{4 m_{F_i}^2}$, while $A_{1/2}^h$ is a loop form factor whose definition is given e.g. in \cite{Djouadi:2005gj}. The matrix $\mathcal{C}_F$ is defined as:
\begin{equation}
\mathcal{C}_F=U_F^L \cdot \mathcal{Y}_F \cdot (U_F^R)^{\dagger},\,\,\,\,\mathcal{Y}_F=\partial_v \mathcal{M}_F=
\frac{1}{\sqrt{2}}
\begin{pmatrix}
0 &
y_h^{F_L} 
\\
y_h^{F_R} 
& 0
\end{pmatrix}.
\end{equation}
\noindent
For a 125 GeV Higgs we can reliably approximate the loop function $A_{1/2}^h (\tau)$ with its asymptotic value, for $A_{1/2}^h (0) = 4/3$, such that the expression~(\ref{SM-App}) simplifies to:
\begin{equation}
\label{eq:ANP_Higgs}
{\cal A}^{h\gamma \gamma}_{\rm VLF}=A_{1/2}^{h} (0) \sum_{F=U,D} \frac{-2 v^2 y_h^{F_L} y_h^{F_R}}{2 M_F M_{UD} - v^2 y_h^{F_L} y_h^{F_R}}.
\end{equation}
\noindent
Experimental measurements do not exhibit statistically relevant deviations of $\mu_{\gamma \gamma}$ from the SM prediction~\cite{Aad:2015gba,Khachatryan:2014jba}, which implies essentially two possibilities: $\mathcal{A}_{\rm VLL}^{h \gamma \gamma} \simeq 0$ or $\mathcal{A}_{\rm VLL}^{h \gamma \gamma} \simeq -2 \mathcal{A}_{\rm SM}^{h \gamma \gamma}$. As evident from eq.~(\ref{eq:ANP_Higgs}), the first possibility is easily realized by setting to zero one of the $y_h^{F_{L,R}}$ couplings.\footnote{Alternatively one could think about a cancellation between the contributions of the ``up'' and ``down'' sectors. In order to have a DM candidate we will consider, however, in this work the case that the up sector is made by electrically neutral states, so that they do not actually contribute to $\mu_{\gamma \gamma}$. On general grounds, a cancellation between the up-type and down-type contribution would be anyway difficult to realize since it would require a very strong deviation from the custodial symmetry limit, which is disfavored by EWPT.} The other is instead more complicated to realize. Assuming $Y=0$ (as will be done for the rest of the paper), such that only $D$-type states contribute to $\mu_{\gamma \gamma}$, and setting for simplicity $M_D=M_{UD}$ and $y_h^{D_L}=-y_h^{D_R}=y_h^D$, which implies that the two mass eigenstates will have the same mass $m_D$, the relation to impose becomes:
\begin{equation}
\mathcal{A}_{\rm VLL}^{h \gamma \gamma}= \frac{4}{3} {\left(\frac{y_h^D v}{m_D}\right)}^2 \simeq -2 \mathcal{A}_{\rm SM}^{h \gamma \gamma} \simeq 13,
\end{equation}
which is impossible to satisfy since $y_h^D v / m_D$ is smaller  than 2 (or equal, for $M_D=M_{UD}=0$).\footnote{This constraint on the Yukawa coupling can be relaxed by adding more families of VLF and/or considering higher values of $Y$. However, we won't consider these cases throughout this work.} Unless differently stated we will always consider, for both the SM and 2HDM cases, an assignation of the Yukawa couplings of the VLFs such that $\mathcal{A}_{\rm SM}^{h\gamma \gamma}=0$.

\subsection{DM Phenomenology}

\noindent
A DM candidate is introduced, in our setup, by considering a ``family'' of vector leptons coupled with the SM Higgs doublet according to the following lagrangian:
\begin{align}
-{\cal L}_{VLL}&= y_h^{N_R} \overline{L}_L \tilde{H} N_R^\prime + y_h^{N_L} \overline{N}^\prime_L \tilde{H}^\dagger L_R +
y_h^{E_R} \overline{L}_L H E^\prime_R + y_h^{E_L} \overline{E}^\prime_L H^\dagger L_R 
\notag \\
&+M_L \overline{L}_L L_R 
+ M_N \overline{N}^\prime_L N^\prime_R +M_E \overline{E}^\prime_L E^\prime_R + \mathrm{h.c.}.
\label{2HDM-VLLs}
\end{align}
To guarantee the stability of the DM candidate, we impose a global $\mathbb{Z}_2$ symmetry under which the vector-like leptons are odd and the SM is even (a supersymmetric analogue is the well-known R-parity).
After EW symmetry breaking a mixing between the vector like fermions is generated, as described by the following mass matrices:

\beq
\mathcal{M}_N = \left( \begin{array}{cc} 
M_N & v' y_h^{N_L} \\ 
v' y_h^{N_R} & M_L 
\end{array}\right), \quad
\mathcal{M}_L = \left( \begin{array}{cc} 
M_E & v' y_h^{E_L} \\ 
v' y_h^{E_R} & M_L 
\end{array}\right).
\label{eq:vll_massmat}
\eeq
where $ v' = v/\sqrt{2} \simeq 174$~GeV. Note that the $\mathbb{Z}_2$ symmetry prevents mixing between the VLLs and the SM fermions. In order to pass from the interaction to the mass basis one has to bidiagonalize the above matrices as:
\beq
U_L^N \cdot \mathcal{M}_N \cdot \left( U_R^N \right)^{\dagger} = \mathrm{diag} (m_{N_1}, m_{N_2}), \quad U_L^E \cdot \mathcal{M}_E \cdot \left( U_R^E \right)^{\dagger} = \mathrm{diag} (m_{E_1}, m_{E_2}),
\eeq
with the unitary matrices $U_{L,R}^{F}$, $F=N,E$ written explicitly as:
\begin{equation}
U_{L,R}^F=\left(
\begin{array}{cc}
\cos \theta_{L,R}^F & \sin \theta_{L,R}^F \nonumber\\
-\sin \theta_{L,R}^F & \cos \theta_{L,R}^F
\end{array}
\right),
\end{equation}

\noindent
where:

\begin{align}
\label{eq:LRangles}
\tan 2 \theta_L^N =\frac{2 \sqrt{2} v \left(M_L y_h^{N_L}+ M_N y_h^{N_R}\right)}{2 M_L^2-2 M_N^2 -v^2 \left(|y_h^{N_L}|^2-|y_h^{N_R}|^2\right)},\nonumber\\
\tan 2 \theta_R^N =\frac{2 \sqrt{2} v \left(M_N y_h^{N_L}+ M_L y_h^{N_R}\right)}{2 M_L^2-2 M_N^2 +v^2 \left(|y_h^{N_L}|^2-|y_h^{N_R}|^2\right)}.
\end{align}
The corresponding expressions for $\theta_{L,R}^E$ can be found from the ones above by replacing $M_N \rightarrow M_E$ and $y_h^{N_{L,R}} \rightarrow y_h^{E_{L,R}}$. 

\noindent
The DM candidate $N_1$ (i.e. the lighter VL neutrino) is in general a mixture of the $SU(2)$ singlet (with null hypercharge) $N_{L,R}^{'}$ and doublet $N_{L,R}$. As a consequence it is coupled with the Higgs scalar $h$ as well as with the SM gauge bosons $W^{\pm}$ and $Z$. These couplings are given by:
\begin{align}
\label{eq:hcoupling}
& y_{h N_1 N_1}= \frac{\cos \theta_N^L \sin \theta_N^R y_h^{N_L}+\cos \theta_N^R \sin \theta_N^L y_h^{N_R}}{\sqrt{2}}, \nonumber\\
& y_{V,Z N_1 N_1}=\frac{g}{4 \cos\theta_W}\left( \sin^2 \theta_L^N + \sin^2 \theta_R^N \right), \nonumber\\
& y_{A,Z N_1 N_1}=\frac{g}{4 \cos\theta_W}\left( \sin^2 \theta_L^N - \sin^2 \theta_R^N \right), \nonumber\\
& y_{V, W N_1 E_1}=\frac{g}{2 \sqrt{2}} \left( \sin \theta_L^N \sin \theta_L^E + \sin \theta_R^N \sin \theta_R^E \right), \nonumber\\
& y_{A, W N_1 E_1}=\frac{g}{2 \sqrt{2}} \left( \sin \theta_L^N \sin \theta_L^E - \sin \theta_R^N \sin \theta_R^E \right),
\end{align} 
where, for convenience, we have expressed the couplings with the $Z$ and $W$ bosons in terms of vectorial and axial combinations.

\noindent
The DM relic density can be determined through the WIMP paradigm as a function of the DM pair annihilation cross-section, being in turn a function of the couplings reported in eq.~\ref{eq:hcoupling}. The possible DM annihilation processes consist into annihilations into SM fermions pairs, induced by s-channel exchange of the $h$ and $Z$ bosons, and into $W^+ W^-$, $ZZ$, $Zh$, and $hh$, induced by t-channel exchange of the neutral states $N_{1,2}$ ($E_{1,2}$ for the $W^+ W^-$ final state). The general expressions of the corresponding cross-sections are rather complicated. We thus provide some schematic expressions~\footnote{These should be just intended as representative expressions to show in a simple way the impact of the relevant parameters of the theory. All the results reported rely on the full numerical evaluation of the cross-sections.} obtained by considering the velocity expansion, i.e. $\langle \sigma v \rangle \approx a + b v^2$ and keeping only the non vanishing contributions for $v \rightarrow 0$.

\noindent
In the case of annihilation into $\bar f f$ final states, the only non vanishing contribution in the $v \rightarrow 0$ limit is the one associatated to the s-channel $Z$-exchange:

\begin{equation}
\langle \sigma v \rangle_{ff} \approx \frac{m_{N_1}^2}{8 \pi} \frac{g^2 m_{N_1}^2}{\pi ((4 m_{N_1}^2-m_Z^2)^2+m_Z^2 \Gamma_Z^2)} 
\sum n_c^f (|V_f|^2+|A_f|^2) |y_{V,Z N_1 N_1}|^2,
\end{equation} 
where $V_f$ and $A_f$ are the vectorial and axial couplings of the $Z$-boson and the SM fermions:
\begin{equation}
V_f=\frac{g}{2 c_W} (-2 q_f s_W^2 +T^3_f),\,\,\,\,\,A_f=\frac{g}{2 c_W} T^3_f,
\end{equation}
while $n_c^f$ is the color factor. The cross-sections of the other relevant final states can be instead estimated as~\footnote{For simplicity we have assumed that the t-channel diagrams are dominated by the exchange of the lightest mass eigenstate.}:

\begin{align}
& \langle \sigma v \rangle_{W^+ W^-} \approx \frac{g^4 t_W}{16 \pi m_W^2} ((\sin \theta_L^N)^2+(\sin \theta_R^N)^2)^2 \nonumber\\
& +\frac{g^4}{64}\left( \frac{1}{2 \pi} ((\sin \theta_L^N \sin \theta_L^E)^2+(\sin \theta_R^N \sin \theta_R^E)^2)^2 \frac{m_{N_1}^2}{(m_{N_1}^2+m_{E_1}^2)^2}\right. \notag \\
& \left.+\frac{2}{\pi}((\sin \theta_L^N \sin \theta_L^E)^2-(\sin \theta_R^N \sin \theta_R^E)^2)^2 \frac{m_{N_1}^4}{m_W^4}  \frac{m_{E_1}^2}{(m_{N_1}^2+m_{E_1}^2)^2}\right),
\end{align}

\begin{align}
& \langle \sigma v \rangle_{ZZ} \approx \frac{g^4}{32 \pi c_W^4 m_Z^2}\left[\frac{m_Z^2}{4 m_{N_1}^2} \left(\left|(\sin \theta_L^N)^2+(\sin \theta_R^N)^2\right|^4\right. \right.\nonumber\\
& \left. \left. +\left|(\sin \theta_L^N)^2-(\sin \theta_R^N)^2\right|^4\right)\right.\nonumber\\
& \left. +2 \left|(\sin \theta_L^N)^2+(\sin \theta_R^N)^2\right|^2 \left|(\sin \theta_L^N)^2-(\sin \theta_R^N)^2\right|^2\right],
\end{align}
and
\begin{equation}
\langle \sigma v \rangle_{Zh} \approx \frac{g^2}{4 \pi v^2} |y_{V, Z N_1 N_1}|^2 \frac{m_Z^2}{m_{N_1}^2}.
\end{equation}

\noindent
The achievement of the correct relic density through DM annihilatations can be potentially in tension with limit from Direct Detection experiments. Indeed, DM interactions with SM quarks, mediated by t-channel exchange of $Z$ and $h$ bosons, induce both Spin Indipendent (SI) and Spin Dependent (SD) scattering processes of the DM with nuclei of target detectors.

\noindent
The corresponding cross-sections, focusing for simplicity on the scattering on protons, are given by: 

\begin{align}
& \sigma_{N_1 p, Z}^{SI}=\frac{\mu_{N_1}^2}{\pi}\frac{1}{m_Z^4}|y_{V,Z N_1 N_1}|^2 
{\left[\left(1+\frac{Z}{A}\right)V_u + \left(2-\frac{Z}{A}\right)V_d\right]}^2\nonumber\\
& \sigma_{N_1 p, h}^{SI}=\frac{\mu_{N_1}^2}{\pi}\frac{m_p^2}{v^2}\left|\frac{y_{h N_1 N_1}}{m_h^2}\left(\sum_{q=u,d,s}f_q +\frac{2}{27}f_{TG} \sum_{q=c,b,t}\right)\right|^2,
\end{align}

\begin{equation}
\sigma_{N_1 p}^{SD}=\frac{3}{m_Z^4} |y_{A, Z N_1 N_1}|^2 \frac{\mu_{N_1}^2}{\pi} \frac{{\left[A_u \left(\Delta_u^p S_p^A+\Delta_u^n S_n^A\right) +A_d \left((\Delta_d^p+\Delta_s^p)S_p^A +(\Delta_d^n+\Delta_s^n)S_n^A\right)\right]}^2}{(S_p^A+S_n^A)^2}
\end{equation}
In the expressions above, $\mu_{N_1}=\frac{m_p m_{N_1}}{m_p+m_{N_1}}$, $f_q, f_{TG},\Delta_q^{p,n}$ are nucleon form factors, while $S_p^A$ and $S_n^A$ are the contributions of the proton and neutron to the spin of the nucleus $A$. We have used the values reported in~\cite{DelNobile:2013sia}.
\noindent
Among these contributions, the most important one is represented by the SI cross-section from $Z$-mediated interactions. This allows to estimate the SI cross-section as:
\begin{equation}
\sigma_{N_1 p}^{\rm SI} \approx 2 \times 10^{-39}\,{\mbox{cm}}^2 {\left( \sin^2 \theta_L^N+\sin^2 \theta_R^N\right)}^2.
\end{equation}
In order to comply with the stringent limits by the LUX experiment~\cite{Akerib:2015rjg} which impose, for DM masses of the order of few hundreds GeV, a cross section of the order of $10^{-45}\,{\mbox{cm}}^2$, we need to require $\sqrt{\sin^2 \theta_L^N+\sin^2 \theta_R^N} \sim 10^{-(1 \div 2)}$. 

\begin{figure}[htb]
\begin{center}
\includegraphics[width=7.5 cm]{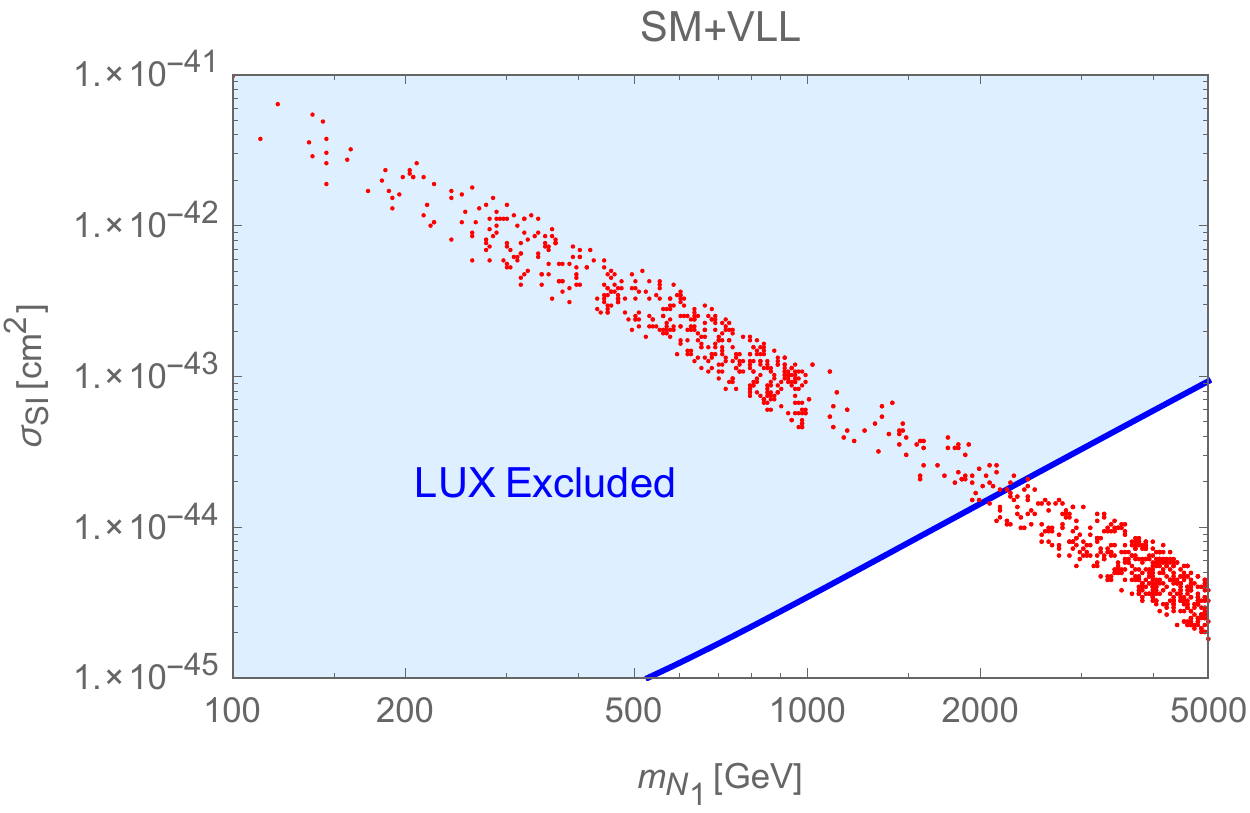}
\end{center}
\caption{Model points satisfying EWPT and Higgs width constraints and providing the correct DM relic density (see main text for clarification) reported in the bidimensional plane $(m_{N_1},\sigma_p^{\rm SI})$. The blue region is excluded by current constraints from DM Direct Detection.}
\label{fig:Higgs_DM}
\end{figure}

\noindent
We have computed the main DM observables, i.e. relic density and SI scattering cross section, for a sample of model points generating by scanning on the parameters $(y_h^{N_{L,R}},y_h^{E_{L}},M_N,M_E,M_L)$, while we set $y_h^{E_R}=0$ in order to achieve $\mathcal{A}_{NP}^{h\gamma \gamma}=0$, over the following range:
\begin{align}
& y_h^{N_{L,R}} \in \left[10^{-3},1\right], \nonumber\\
& y_h^{E_{L}} \in \left[5 \times 10^{-3},3\right], \nonumber\\
& M_N \in \left[100\,\mbox{GeV}, 5\,\mbox{TeV}\right], \nonumber\\
& M_E=M_L \in \left[300\,\mbox{GeV}, 5\,\mbox{TeV}\right],
\end{align}
with the additional requirement of not exceeding the limits from EWPT. 

\noindent
The results of our analysis are reported in fig.~\ref{fig:Higgs_DM}. The figure shows the set of points featuring the correct DM relic density in the bidimensional plane $(m_{N_1},\sigma_{\rm SI})$. As evident, the very strong constraints from the $Z$-mediated DM scattering on nucleons rule out the parameter space corresponding to thermal Dark Matter unless its mass is approximately above 2~TeV. This result is very similar to what is obtained in the generic scenario dubbed $Z$-portal~\cite{Arcadi:2014lta}, in which the SM $Z$ boson mediates the interactions between the SM states and a Dirac fermion DM candidate. Notice that in our parameter scan we have anyway imposed the existence of a sizable mass splitting between the DM and the ligthest electrically charged fermion $E_1$. If this constraint were relaxed, DM annihilations would be enhanced by coannihilation effects and the tension with Direct Detection limits would be possibly relaxed (see e.g.~\cite{Yaguna:2015mva}).

%%%%%%%%%%%%%%%%%%%%%%%%%%%%%%%%%%%%%%%%%%%%%%%%%%%%%%%%%%%%%%%%%%%%%%%%%%
\section{Two Higgs Doublet Models}
\label{2HDM}
%%%%%%%%%%%%%%%%%%%%%%%%%%%%%%%%%%%%%%%%%%%%%%%%%%%%%%%%%%%%%%%%%%%%%%%%%%

\noindent
Let us now move to the case of 2HDM scenarios. We will summarize below the most salient features of this scenario and fix as well the notation. For a more extensive review we refer instead, for example, to \cite{Branco:2011iw}.

\begin{figure}[!t]
%\vspace*{-7mm}
\begin{center}
\includegraphics[keepaspectratio=true,width=0.4\textwidth]{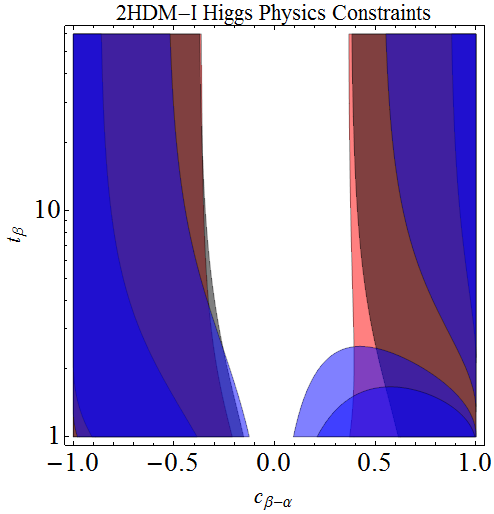}
\quad
\includegraphics[keepaspectratio=true,width=0.4\textwidth]{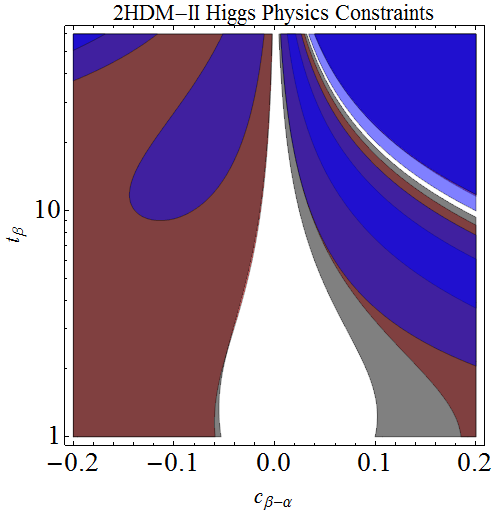}
\\
\includegraphics[keepaspectratio=true,width=0.4\textwidth]{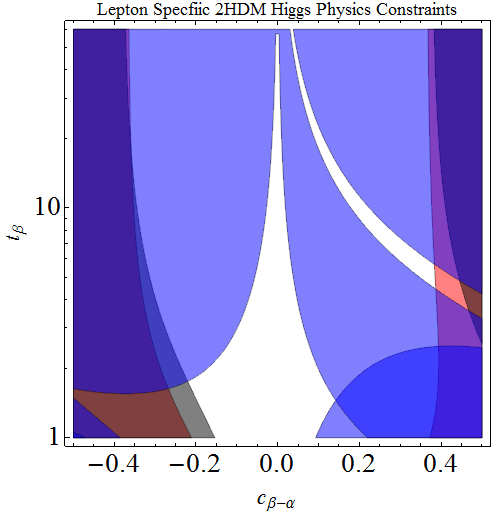}
\quad
\includegraphics[keepaspectratio=true,width=0.4\textwidth]{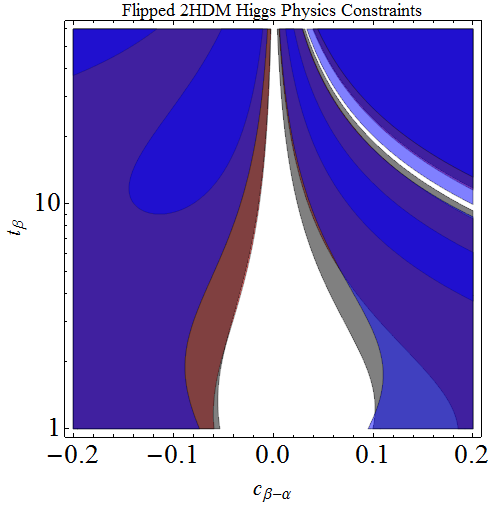}
\caption{
{\small Constraints in the $\left(c_{\beta-\alpha}, t_{\beta}\right)$ plane on the four types of flavour-conserving 2HDMs, coming from Higgs signal strength measurements~\cite{Aad:2015gba,Khachatryan:2014jba}. The signal strengths we have considered are $\mu_{\gamma\gamma}$ (red), $\mu_{ZZ,WW}$ (grey), and $\mu_{bb,\tau\tau}$ (blue).}  
}
\label{Fig:Higgs_constr}
\end{center}
%\vspace*{-7mm}
\end{figure}

\noindent
The scalar potential of the (CP-conserving) 2HDM is given by:
\begin{align}
\label{eq:scalar_potential}
V(H_1,H_2) &= 
m_{11}^2 H_1^\dagger H_1+ m_{22}^2 H_2^\dagger H_2
- m_{12}^2 \left(H_1^\dagger H_2 + {\rm h.c.} \right)
+\frac{\lambda_1}{2} \left( H_1^\dagger H_1 \right)^2
+\frac{\lambda_2}{2} \left( H_2^\dagger H_2 \right)^2
\notag
\\
&+ \lambda_3 \left(H_1^\dagger H_1 \right)\left(H_2^\dagger H_2 \right)
+ \lambda_4 \left(H_1^\dagger H_2 \right)\left(H_2^\dagger H_1 \right)
+\frac{\lambda_5}{2}\left[ \left(H_1^\dagger H_2 \right)^2 + {\rm h.c.} \right],
\end{align}
where two doublets are defined by:
\begin{equation}
H_i=
\begin{pmatrix}
\phi_i^+ \\
(v_i+\rho_i +i \eta_i)/\sqrt{2}
\end{pmatrix}~,
\qquad
i=1,2,
\end{equation}
where, as usual, $v_2/v_1= \tan \beta \equiv t_{\beta}$. The spectrum of physical states is constituted by two CP even neutral states, $h$, identified with the 125 GeV Higgs, and $H$, the CP-odd Higgs $A$ and finally the charged Higgs $H^{\pm}$. The transition from the interaction basis $(H_1,H_2)^T$ to the mass basis $(h,H,A,H^{\pm})$ depends on two mixing angles, $\alpha$ and $\beta$. Throughout all this work we will assume to be in the so called alignment limit, i.e. $\alpha\simeq\beta-\pi/2$. This is a reasonable assumption since, in most scenarios, as also shown in fig.~\ref{Fig:Higgs_constr}, only small deviations from the alignment limit are experimentally allowed. In this limit, the $h$ boson becomes completely SM-like. A second relevant implication is that the couplings of the second CP Higgs $H$ with $W$ and $Z$ bosons are zero at tree level, being proportional to $\cos(\beta-\alpha)$ (analogous tree-level couplings for the $A$ boson are forbidden by CP conservation). For a more detailed treatment of the alignment limit, we refer the reader to e.g. Refs.~\cite{Gunion:2002zf,Carena:2013ooa,Wang:2013sha,Bernon:2015qea}.

\noindent
The quartic couplings of the scalar potential~(\ref{eq:scalar_potential}) can be expressed as function of the masses the physical states as:
\begin{align}
\label{eq:quartic_physical}
\lambda_1 &= \frac{1}{v^2} \left[ m_h^2 + \left( m_H^2 - M^2 \right) t_{\beta}^2 \right], \\
\lambda_2 &= \frac{1}{v^2} \left[ m_h^2 + \left( m_H^2 - M^2 \right) t_{\beta}^{-2} \right], \\
\lambda_3 &= \frac{1}{v^2} \left[ m_h^2 + 2 m_{H^{\pm}}^2 - \left( m_H^2 + M^2 \right) \right], \\
\lambda_4 &= \frac{1}{v^2} \left[ M^2 + m_A^2 - 2 m_{H^{\pm}} \right], \\
\lambda_5 &= \frac{1}{v^2} \left[ M^2 - m_A^2 \right], 
\end{align} 
where $M \equiv m_{12}/(s_{\beta} c_{\beta})$. Unitarity and boundedness from below of the scalar potential impose constraints on the value of the couplings $\lambda_{i=1,5}$~\cite{Branco:2011iw,Becirevic:2015fmu} which, through eq.~\ref{eq:quartic_physical}, are translated into bounds on the physical masses. In particular these bounds imply that it is not possible to assign their values independently one from each other. 
All these bounds can be found, for example, in Ref.~\cite{Kanemura:2004mg,Becirevic:2015fmu}, but, for completeness, we will report them below. For the scalar potential to be bounded from below, the quartics must satisfy:
\begin{equation}
\label{eq:up1}
\lambda_{1,2} > 0, \; \lambda_3 > -\sqrt{\lambda_1\lambda_2}, \; {\rm and} \; \lambda_3 + \lambda_4 - \left|\lambda_5\right| > -\sqrt{\lambda_1\lambda_2},
\end{equation}
while s-wave tree level unitarity imposes that:
\beq
\label{eq:up2}
\left| a_{\pm} \right|, \left| b_{\pm} \right|, \left| c_{\pm} \right|, \left| f_{\pm} \right|, \left| e_{1,2} \right|, \left| f_1 \right|, \left| p_1 \right| < 8\pi,
\eeq
where:
\begin{align}
a_{\pm} &= \frac{3}{2}(\lambda_1 + \lambda_2) \pm \sqrt{\frac{9}{4}(\lambda_1-\lambda_2)^2 + (2\lambda_3 + \lambda_4)^2}, \notag \\
b_{\pm} &= \frac{1}{2}(\lambda_1 + \lambda_2) \pm \sqrt{(\lambda_1-\lambda_2)^2 + 4\lambda_4^2}, \notag \\
c_{\pm} &= \frac{1}{2}(\lambda_1 + \lambda_2) \pm \sqrt{(\lambda_1-\lambda_2)^2 + 4\lambda_5^2}, \notag \\
e_1 &= \lambda_3 + 2\lambda_4 - 3\lambda_5, \quad e_2 = \lambda_3 - \lambda_5, \notag \\
f_+ &= \lambda_3 + 2\lambda_4 + 3\lambda_5, \!\quad f_- = \lambda_3 + \lambda_5, \notag \\
f_1 &= \lambda_3 + \lambda_4, \quad p_1 = \lambda_3 - \lambda_4.
\end{align}
Later on, we will include these constraints as well when doing our scans.
\noindent
Motivated by the non-observation of flavour-changing neutral currents (FCNCs), we choose to couple the SM fermions to only one scalar doublet~\cite{Branco:2011iw}. Consequently, the couplings of the SM fermions to the Higgses are:
\begin{align}
-{\cal L}_{yuk}^{SM} &=\sum\limits_{f=u,d,l} \frac{m_f}{v} \left[\xi^f_h \overline{f}f h+\xi^f_H \overline{f}f H-i \xi^f_A \overline{f}\gamma_5 f A \right]
\notag
\\
&-\left[\frac{\sqrt{2}}{v} \overline{u} \left(m_u \xi^u_A P_L + m_d \xi^d_A P_R \right)d H^+ +
\frac{\sqrt{2}}{v} m_l \xi_A^l \overline{\nu_L} l_R H^+  + \mathrm{h.c.} \right],
\end{align}
where the $\xi$'s for the four flavour-conserving types of 2HDMs are listed below in table~\ref{table:2hdm_type}. On the contrary, we couple the VL fermions to both doublets:\footnote{
Since we are coupling the VLFs to both doublets, we cannot rigorously refer to type-I, type-II, Lepton-Specific or Flipped 2HDMs, as flavor violating yukawa couplings, possibly responsible for FCNCs, might be induced radiatively by the VLLs. We will nevertheless retain 
the classification of the various 2HDM realizations in order to distinguish the different dependence on $\tan\beta$ of the couplings of the SM fermions and the Higgs mass eigenstates.}
\begin{align}
\label{2HDM_VL_Lag}
-{\cal L}_{\rm VLL} &= y_i^{U_R} \overline{{\cal D}_L} \tilde{H}_i U^\prime_R + y^{U_L}_i \overline{U^\prime_L} \tilde{H}_i^\dagger {\cal D}_R +
y^{D_R}_i \overline{{\cal D}_L} H_ i D^\prime_R + y^{D_L}_i \overline{D^\prime_L} H_ i^\dagger {\cal D}_R 
\notag \\
&+M_{\mathcal{D}} \overline{{\cal D}_L} {\cal D}_R 
+ M_U \overline{U^\prime_L} U^\prime_R +M_D \overline{D^\prime_L} D^\prime_R + {\rm h.c.},
\end{align}
where a sum over $i=1,2$ is implied. It is possible to define the Yukawa couplings, $y_h^X$ and $y_H^X$, to the physical CP-even states through the following rotations: 
\beq
\begin{pmatrix} y_h^X \\ y_H^X \end{pmatrix} = 
\begin{pmatrix} c_{\beta} & s_{\beta} \\ s_{\beta} & -c_{\beta} \end{pmatrix} \begin{pmatrix} y_1^X \\ y_2^X \end{pmatrix}, \quad \begin{pmatrix} H_{\rm SM} \\ H_{\rm NP} \end{pmatrix} = 
\begin{pmatrix} c_{\beta} & s_{\beta} \\ s_{\beta} & -c_{\beta} \end{pmatrix} \begin{pmatrix} H_1 \\ H_2 \end{pmatrix},
\label{eq:Higgs_basis}
\eeq
where we used the superscript $X = U_{L/R}$ or $D_{L/R}$. As we are working in the alignment limit, $H_{\rm SM}$ becomes the SM Higgs double, while $H_{\rm NP} = \begin{pmatrix} H^+ \\ (H-iA)/\sqrt{2} \end{pmatrix}$. Since we are coupling the VL fermions to both doublets, the value of $t_{\beta}$ or the chosen type of 2HDM will be irrelevant for the VLF coupling to the scalars. On the contrary, the Yukawa couplings of the SM fermions are dictated exactly by the choices of $t_{\beta}$ and of the 2HDM type.

\begin{table}[h!]
\renewcommand{\arraystretch}{1.3}
\begin{center}
\begin{tabular}{|c|c|c|c|c|}
\hline
 &  Type I & Type II & Lepton-specific & Flipped \\
\hline\hline 
$\xi_h^u$ & $c_{\alpha}/ s_{\beta} \rightarrow 1$ & $c_{\alpha}/ s_{\beta} \rightarrow 1$ & $c_{\alpha}/ s_{\beta} \rightarrow 1$ & $c_{\alpha}/ s_{\beta}\rightarrow 1$
\\
\hline
$\xi_h^d$ & $c_{\alpha}/ s_{\beta} \rightarrow 1$ & $-s_{\alpha}/ c_{\beta} \rightarrow 1$ & $c_{\alpha}/ s_{\beta} \rightarrow 1$ & $-s_{\alpha}/ c_{\beta} \rightarrow 1$
\\
\hline
$\xi_h^l$ & $c_{\alpha}/ s_{\beta} \rightarrow 1$ & $-s_{\alpha}/ c_{\beta} \rightarrow 1$ & $-s_{\alpha}/ c_{\beta} \rightarrow 1$ & $c_{\alpha}/ s_{\beta} \rightarrow 1$ 
\\
\hline\hline
$\xi_H^u$ & $s_{\alpha}/ s_{\beta} \rightarrow -t_{\beta}^{-1}$ & $s_{\alpha}/ s_{\beta} \rightarrow -t_{\beta}^{-1}$ & $s_{\alpha}/ s_{\beta} \rightarrow -t_{\beta}^{-1}$ & $s_{\alpha}/ s_{\beta} \rightarrow -t_{\beta}^{-1}$
\\
\hline
$\xi_H^d$ & $s_{\alpha}/ s_{\beta} \rightarrow -t_{\beta}^{-1}$ & $c_{\alpha}/ c_{\beta} \rightarrow t_{\beta}$ & $s_{\alpha}/ s_{\beta} \rightarrow -t_{\beta}^{-1}$ & $c_{\alpha}/ c_{\beta} \rightarrow t_{\beta}$
\\
\hline
$\xi_H^l$ & $s_{\alpha}/ s_{\beta} \rightarrow -t_{\beta}^{-1}$ & $c_{\alpha}/ c_{\beta} \rightarrow t_{\beta}$ & $c_{\alpha}/ c_{\beta} \rightarrow t_{\beta}$ & $s_{\alpha}/ s_{\beta} \rightarrow -t_{\beta}^{-1}$
\\
\hline\hline
$\xi_A^u$ & $t_{\beta}^{-1}$ & $t_{\beta}^{-1}$ & $t_{\beta}^{-1}$ & $t_{\beta}^{-1}$
\\
\hline
$\xi_A^d$ & $-t_{\beta}^{-1}$ & $t_{\beta}$ & $-t_{\beta}^{-1}$ & $t_{\beta}$
\\
\hline
$\xi_A^l$ & $-t_{\beta}^{-1}$ & $t_{\beta}$ & $t_{\beta}$ & $-t_{\beta}^{-1}$
\\
\hline
\end{tabular}
\end{center}
\caption{Couplings of the Higgses to the SM fermions as a function of the angles $\alpha$ and $\beta$ and in the alignment limit where $(\beta-\alpha) \rightarrow \pi/2$.}
\label{table:2hdm_type}
\end{table}

\noindent
A DM candidate is again straightforwardly introduced by considering a lagrangian of the form~(\ref{2HDM_VL_Lag}) with $U \equiv N$ and $D \equiv E$. Our analysis will substantially follow the same lines as in the case of VLL extensions of the SM Higgs sector. Before determining the DM observables and comparing them with experimental constraints, we will reformulate, in the next subsections, for the case of the 2HDM, the constraints from the SM Higgs signal strength and from EWPT. We will also consider an additional set of constraints, which influence the size of the new Yukawa couplings, from the UV behavior of the theory.

\subsection{Higgs Signal Strengths}

\noindent
Having imposed an alignment limit, the extended Higgs sector does not influence the decay branching fractions of the 125~GeV SM-like Higgs. The only possible source of deviation from the SM expectation is represented by the VLLs, which can affect the $h\to\gamma\gamma$ signal strength, $\mu_{\gamma\gamma}$. The corresponding contribution substantially coincides with the one determined in the one Higgs doublet scenario, namely eq.~(\ref{eq:ANP_Higgs}). Assuming the presence of only one family of VLLs, the simplest solution for having an experimentally viable scenario is to set to zero one of the $y_h^{E_{L,R}}$ couplings. Unless differently stated, we will assume, in the analysis below, that $y_h^{E_R}=0$.

\subsection{EWPT Constraints}

\noindent
In a 2HDM+VLL framework new contributions, with respect to the SM, to the $S$ and $T$ parameters originate from both the fermionic and the scalar sector. For what regards the former,  these contributions depend, as for the case of one Higgs doublet, on the masses of the new fermions and their couplings $y_h^{N_{L,R},E_{L,R}}$ with the SM-like Higgs, while the couplings with the other Higgs states are unconstrained by EWPT. The contributions from the scalar sector are instead related to the masses of the new Higgs states. Also in this case is possible to forbid deviations from the SM expectations of the $T$ parameter by imposing a custodial symmetry. In the alignment limit this is realized by setting $m_H \simeq m_{H^\pm}$ or $m_A \simeq m_{H^\pm}$~\cite{Davidson:2010xv,Baak:2011ze} and consider only constraints from the $S$ parameter. As already pointed and further clarified below, this choice would imply excessive limitations to DM phenomenology. For this reason we will not impose a custodial symmetry neither to the fermionic nor to the scalar sector, but rather freely vary the corresponding parameters and require in turn that the $S$ $T$ parameters do not deviate by more than $3\sigma$ from their best fit values.

\begin{figure}[htb]
\begin{center}
\includegraphics[width=7 cm]{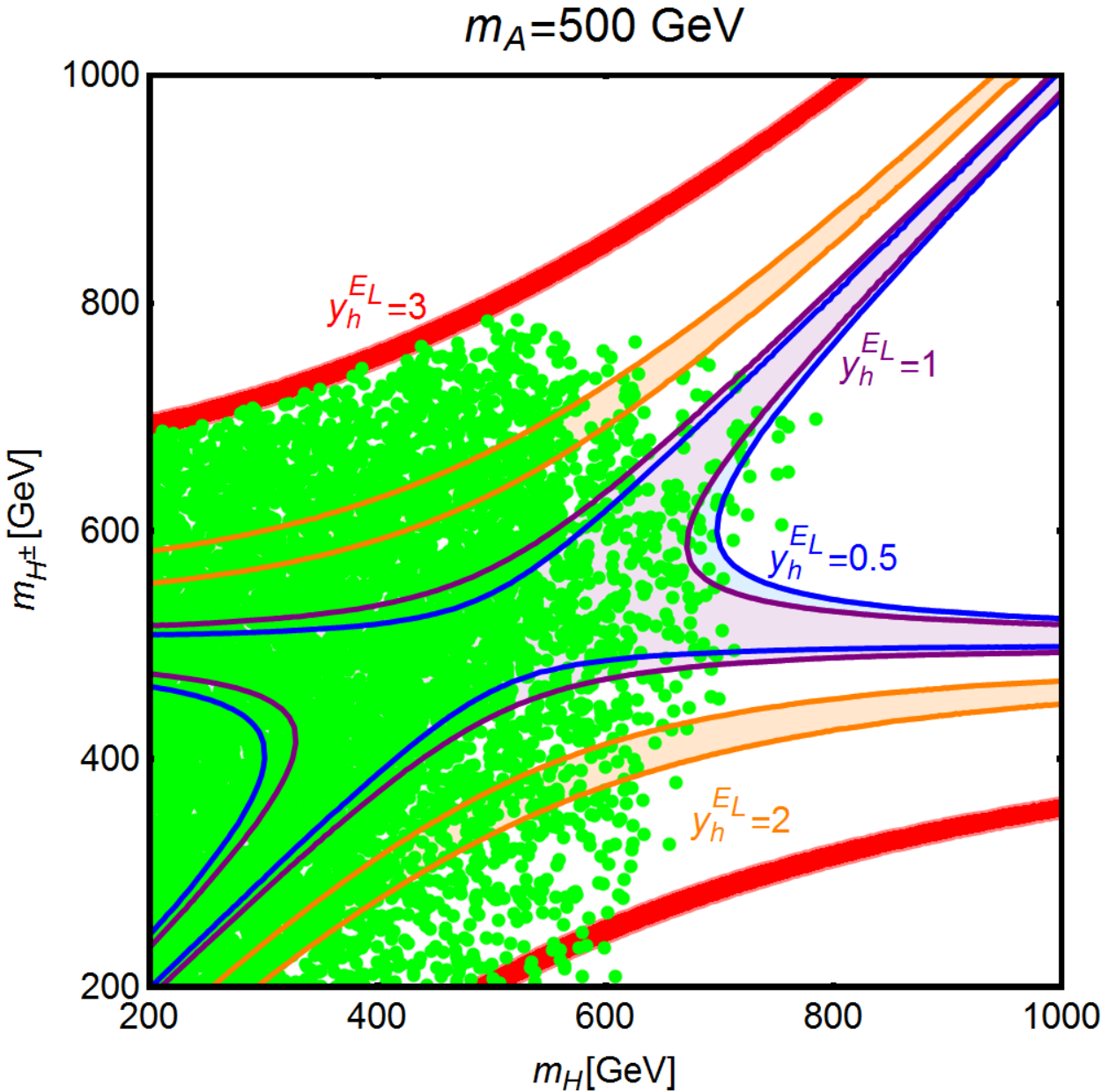}
\hspace{3 mm}
\includegraphics[width=7 cm]{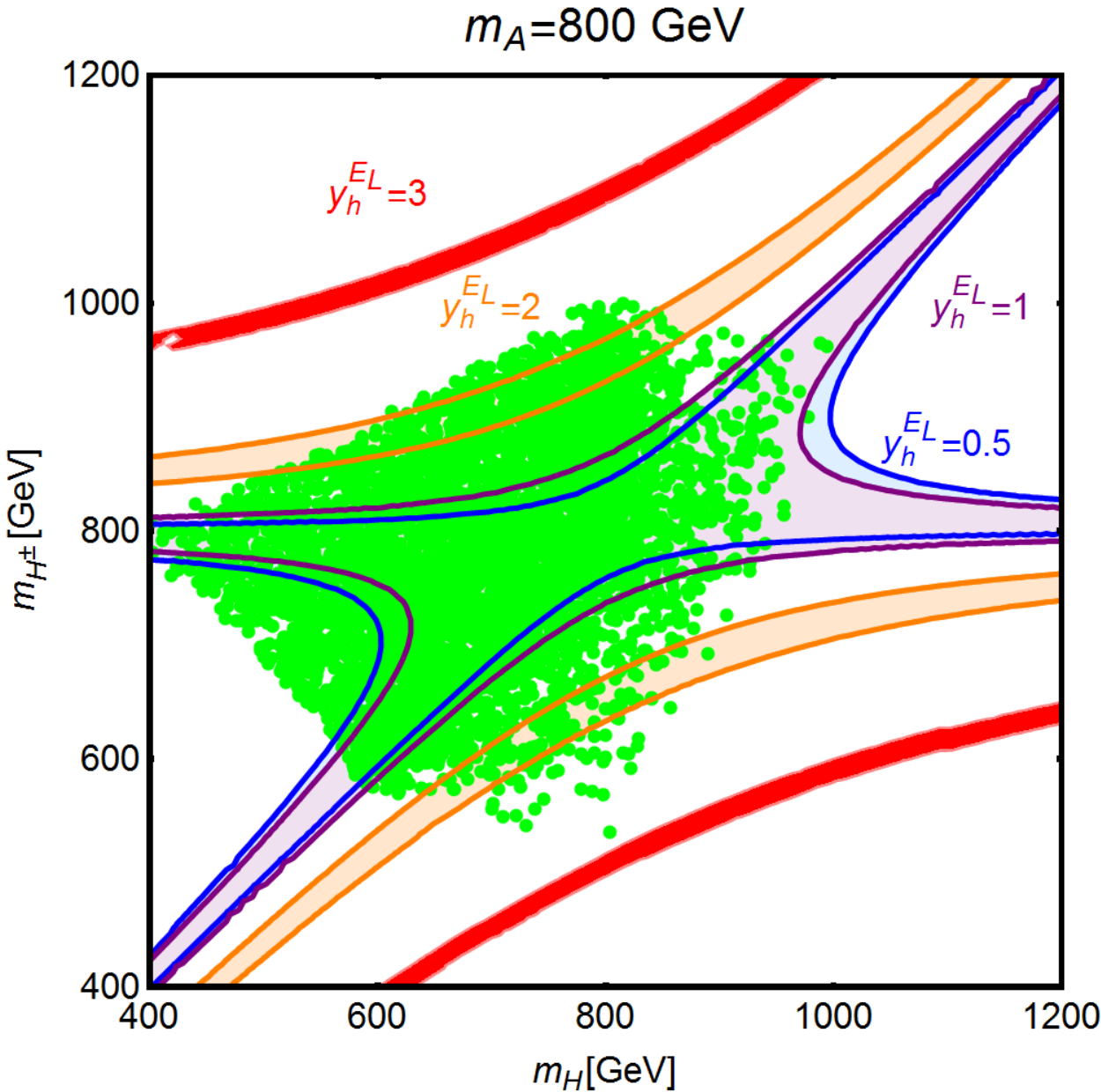}
\end{center}
\caption{Impact of EWPT constraints in the bidimensional plane $(m_H,m_{H^{\pm}})$ for two fixed assignations of $m_A$, i.e. 500 and  800 GeV. The blue,purple,orange and red regions represent the allowed parameter space for, respectively, $y_h^{E_L}=0.5,1,2,3$. The green points represent the configurations allowed by the constraints reported in eq.~(\ref{eq:up1}) and~(\ref{eq:up2}).}
\label{fig:EWPT_2HDM}
\end{figure}

\noindent
For illustrative purposes we have reported in fig.~(\ref{fig:EWPT_2HDM}) the regions allowed by EWPT for some definite assignation of model parameters. More specifically we have fixed the values of the DM candidate $M_{N_1}$ and of the lightest charged new fermion $m_{E_1}$, as well as the Yukawa coupling $y_h^{N_L}$, to, respectively, 120 GeV, 250 GeV and 0.01 (this very low value is motivated by constraints from DM DD), while we have varied the parameter $y_h^{E_L}$, since it will be relevant for the DM relic density as well as for LHC detection prospects. Regarding the scalar sector we have fixed $m_A=500$~GeV (left panel) and $m_A=800$~GeV (right panel) and varied the mass of the CP-even Higgs state $H$ and of the charged one $H^{\pm}$. For $y_h^{E_L} \leq 1$ the effect of the fermionic sector on the EWPT is subdominant such that the allowed regions substantially corresponds to the one allowed in the case of no VLLs present in the theory. On the contrary, once the value of $y_h^{E_L}$ is increased, a cancellation between the contributions from the fermionic and scalar sectors is needed in order to comply with experimental constraints. As consequence the allowed regions of the parameter space are reduced to rather narrow bands. We also notice that, in this last case, the constraints from EWPT disfavor mass degenerate $H,A,H^{\pm}$. We remind that, on the other hand, the variation of the masses of the Higgs states is constrained by perturbativity and unitarity limits, eq.(\ref{eq:up1})-(\ref{eq:up2}). We have then reported on fig.~(\ref{fig:EWPT_2HDM}) the allowed regions, by these latter constraints, determined by varying the input parameters of eq.~(\ref{eq:quartic_physical}) over the same ranges considered in~\cite{Becirevic:2015fmu} (contrary to this reference we have nevertheless assumed alignment limit). As we can see, values of $y_h^{E_L}$ above 3 are excluded for $m_A=500\,\mbox{GeV}$ while for $m_A=800\,\mbox{GeV}$ we get the even stronger constraint $y_h^{E_L} \lesssim 2$.

%%%%%%%%%%%%%%%%%%%%%%%%%%%%%%%%%%%%%%%%%%%%%%%%%%%%%%%%%%%%%%%%%%%%%%%%%%
\subsection{Constraints from RGE Evolution}
\label{Coupling-evolution}
%%%%%%%%%%%%%%%%%%%%%%%%%%%%%%%%%%%%%%%%%%%%%%%%%%%%%%%%%%%%%%%%%%%%%%%%%%

\noindent
The extension of the Higgs sector with VLFs suffers also constraints from theoretical consistency. Indeed, the presence of new fermions affects the RGE evolution of the parameters of the 2HDM, in particular the gauge couplings and the quartic couplings of the scalar potential~\cite{Bertuzzo:2016fmv}, making it difficult for the new states to induce sizable collider signals, like diphoton events~\cite{Franceschini:2015kwy,Gu:2015lxj,Son:2015vfl,Salvio:2015jgu,Salvio:2016hnf,Bae:2016xni,Hamada:2016vwk,Arcadi:2016acg} (see also below).

\noindent
For what regards the gauge couplings, their $\beta$ functions receive a positive contribution depending on the number of families of vector-like fermions and on their quantum numbers under the SM model gauge group. In case that these contributions are too high the gauge couplings can be lead to a Landau pole at even moderate/low energy scales. However, in the case considered in this work, i.e. one family of vector like leptons, we have only a small contribution to the $\beta$ functions of the couplings $g_1$ and $g_2$ which does not affect in a dangerous way their evolution with energy.

\noindent
Very different is, instead, the case of the quartic couplings. The radiative corrections associated to the VLLs depend on their Yukawa couplings. The $\beta$ functions are, indeed, given by: 
\begin{align}
\label{eq:lambda1RGE}
& \beta_{\lambda_1}=\beta_{\lambda_1,\rm 2HDM}+\frac{1}{8 \pi^2}\left(\lambda_1 \sum_L |y_1^L|^2
%+\lambda_7 \sum_L y_1^L y_2^L
-\sum_L |y_1^L|^4\right),
\end{align}
\begin{align}
\label{eq:lambda2RGE}
& \beta_{\lambda_2}=\beta_{\lambda_2,\rm 2HDM}+\frac{1}{8 \pi^2}\left(\lambda_2 \sum_L |y_2^L|^2
%+\lambda_6 \sum_L y_1^L y_2^L
- \sum_L |y_2^L|^4\right),
\end{align}

\begin{align}
\label{eq:lambda3RGE}
& \beta_{\lambda_3}=\beta_{\lambda_3, \rm 2HDM}+\frac{1}{16 \pi^2} \left(\lambda_3 \sum_L (|y_1^L|^2+|y_2^L|^2)\right. \nonumber\\ 
%+(\lambda_6+\lambda_7) \sum_L y_1^L y_2^L\right.\nonumber\\
&\left.-2 y_1^{E_L} y_2^{E_L} y_1^{N_L}y_2^{N_L}+(|y_1^{N_L}|^2+|y_1^{E_L}|^2)(|y_2^{N_L}|^2+|y_2^{E_L}|^2)\right.\nonumber\\
& \left. -2 y_1^{E_R} y_2^{E_R} y_1^{N_R}y_2^{N_R}+(|y_1^{N_R}|^2+|y_1^{E_R}|^2)(|y_2^{N_R}|^2+|y_2^{E_R}|^2)\right),
\end{align}

\begin{align}
\label{eq:lambda4RGE}
& \beta_{\lambda_4}=\beta_{\lambda_4, \rm 2HDM}+\frac{1}{16 \pi^2} \left(\lambda_4 \sum_L (|y_1^L|^2+|y_2^L|^2)\right. \nonumber\\
%+(\lambda_6+\lambda_7) \sum_L y_1^L y_2^L\right.\nonumber\\
&\left. -2 y_1^{E_L} y_2^{E_L} y_1^{N_L}y_2^{N_L}+(|y_1^{N_L}|^2-|y_1^{E_L}|^2)(|y_2^{N_L}|^2-|y_2^{E_L}|^2)\right.\nonumber\\
&\left. +2 y_1^{E_R} y_2^{E_R} y_1^{N_R}y_2^{N_R}+(|y_1^{N_R}|^2-|y_1^{E_R}|^2)(|y_2^{N_R}|^2-|y_2^{E_R}|^2)\right),
\end{align}

\begin{align} 
\label{eq:lambda5RGE}
& \beta_{\lambda_5}=\beta_{\lambda_5,\rm 2HDM}+\frac{1}{16 \pi^2} \left( \lambda_4 \sum_L (|y_1^L|^2+|y_2^L|^2)
%+ (\lambda_6+\lambda_7) \sum_L y_1^L y_2^L
-2 \sum_L |y_1^L|^2 |y_2^L|^2\right),
\end{align}

\noindent
where $\beta_{\lambda_i,\rm 2HDM}$ are the contributions to the $\beta$ function originating only from the quartic couplings themselves and the Yukawa couplings of the SM fermions. We refer to~\cite{Branco:2011iw} for their explicit expressions.

\noindent
To simplify the notation we have expressed, in eq.\ref{eq:lambda1RGE}-\ref{eq:lambda5RGE}~\footnote{Notice that even if the couplings $\lambda_6$ and $\lambda_7$ have been set to zero, they are radiatively generated. So one should also consider their $\beta$ function as well as additional terms in eq.\ref{eq:lambda1RGE}-\ref{eq:lambda5RGE}. For simplicity we have not explicitly reported these contributions but we have included them in our numerical computations.}, the Yukawa couplings in the $(H_1,H_2)$ basis.

\noindent
As evident, the quartic couplings receive large radiative corrections scaling either with the second or the fourth power of the Yukawa couplings. As a consequence, vacuum stability and/or perturbativity and unitarity might be spoiled at some given energy scale unless additional degrees of freedom are introduced in the theory.

\noindent
A quantitative analysis would require the solution of eq.\ref{eq:lambda1RGE}-\ref{eq:lambda5RGE} coupled with RGE for the gauge and Yukawa couplings as function of the masses of the Higgs eigenstates and the parameters $M$ and $t_\beta$, which determine the initial conditions for $\lambda_{1,5}$, and verify conditions~\ref{eq:up1} and~\ref{eq:up2} as function of the energy scale. A good qualitative understanding can be nevertheless achieved by noticing that for sizable Yukawa couplings the $\beta$ functions~\ref{eq:lambda1RGE}-\ref{eq:lambda5RGE} are dominated by the negative contributions scaling with the fourth power of the Yukawas themselves (their $\beta$ function are positive, scaling qualitatively as $\beta_y \propto y^3$). As a consequence one can focus, among~\ref{eq:up1} and~\ref{eq:up2}, on the vacuum stability conditions $\lambda_{1,2}>0$. A given set of model parameters can be regarded as (at least phenomenologically) viable if the scale at which the couplings become negative is far enough from the one probed by collider processes. In order to have this, the quartic couplings $\lambda_1$ and $\lambda_2$ should not vary too fast with the energy. As proposed in~\cite{Goertz:2015nkp}, a good approximate condition consists in imposing $|\beta_{\lambda_{1,2}}/\lambda_{1,2}| < 1$, with $\lambda_{1,5}$ computed according to eq.~(\ref{eq:quartic_physical}) and the Yukawa couplings set to their input value at the EW scale. In case this condition is not fulfilled, the functions $\beta_{\lambda_{1,2}}$ would vary too fast with the energy so that the theory would manifest a pathological behavior already in proximity of the energy threshold corresponding to the masses of the VLLs~\footnote{We remark that our discussion should be intended as rather qualitative since it is based on 1-loop $\beta$-functions.}.  

\begin{figure}[htb]
\begin{center}
\includegraphics[width=7 cm]{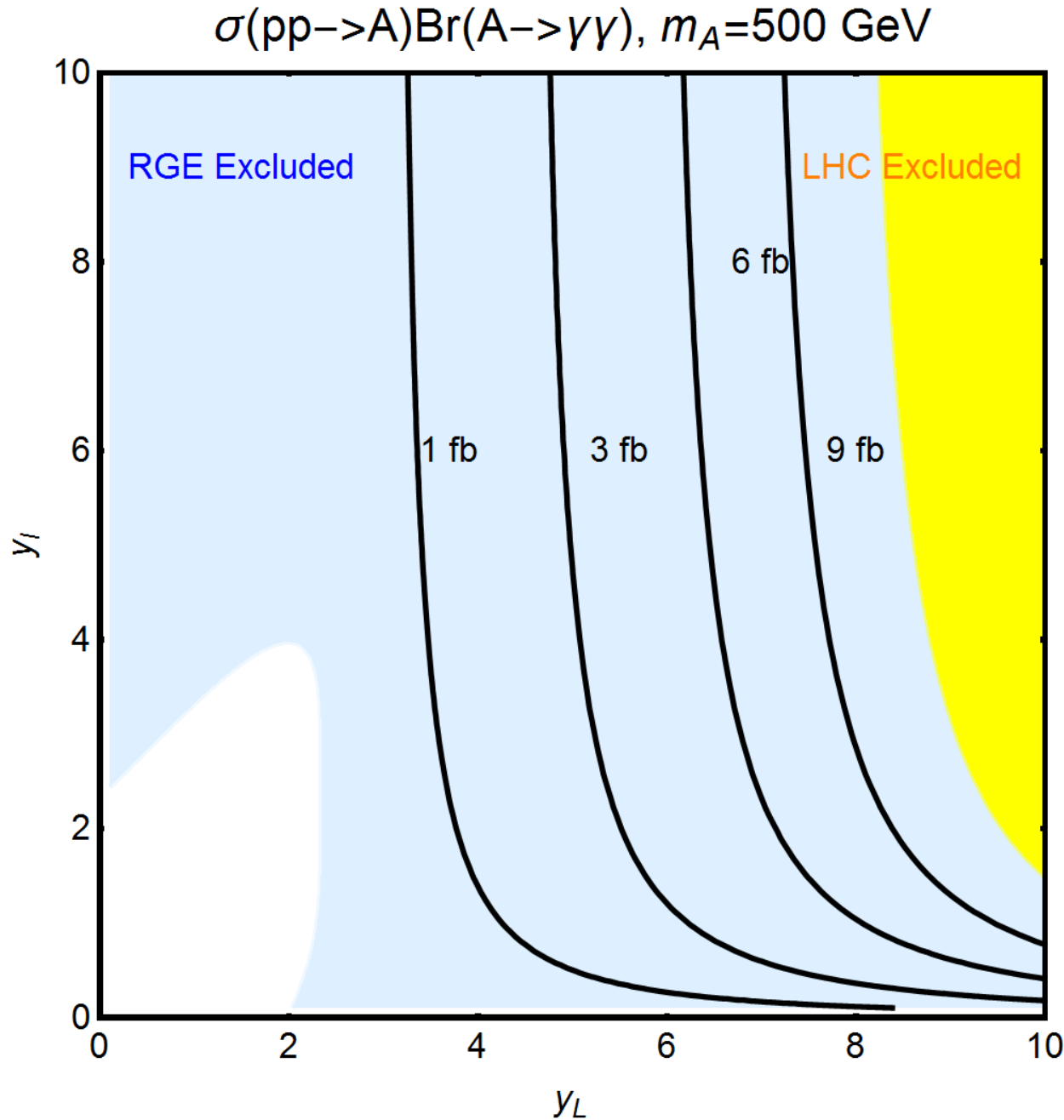}
\hspace{3 mm}
\includegraphics[width= 7 cm]{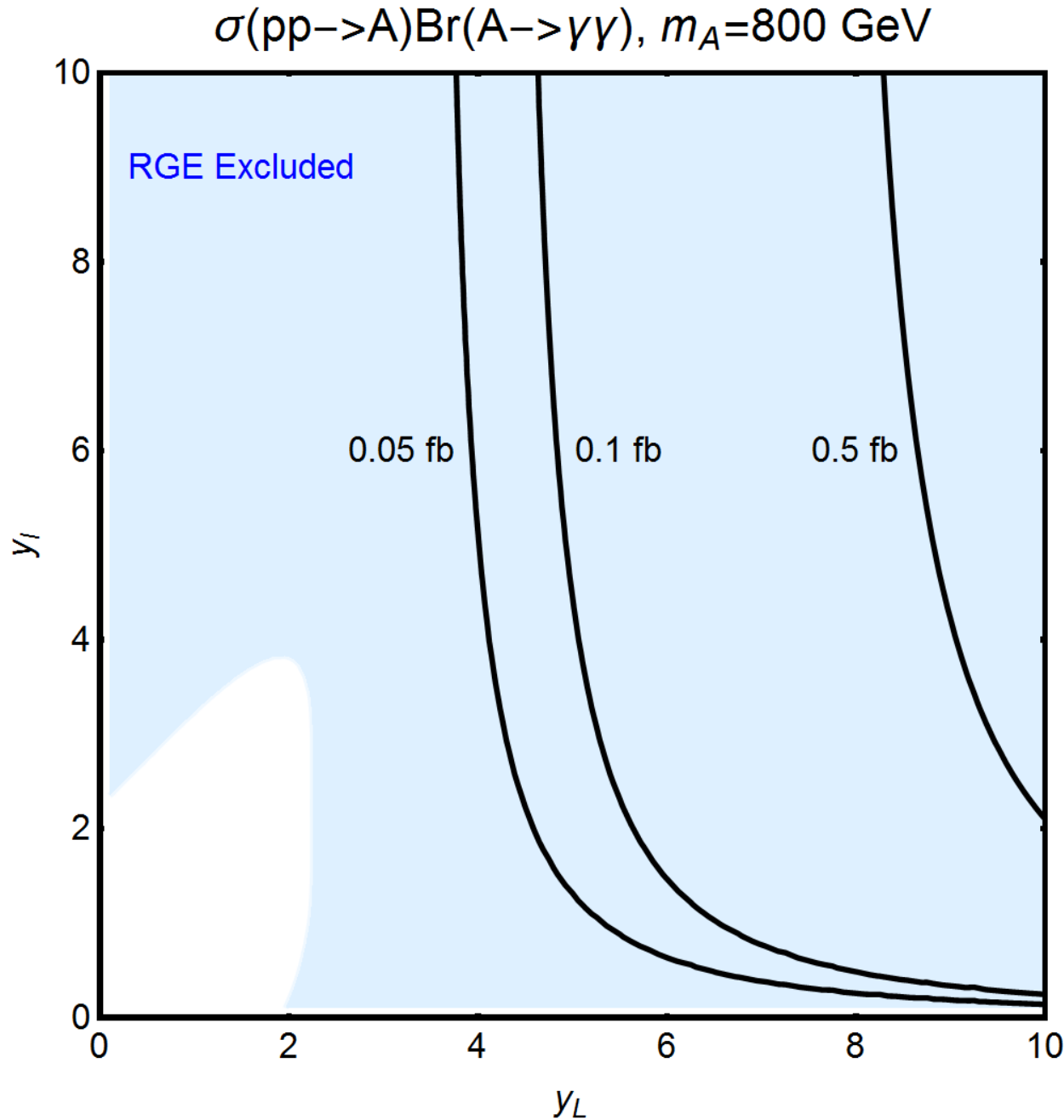}
\end{center}
\caption{Contours of the process $pp \rightarrow A \rightarrow \gamma \gamma$ for the two values $m_A=500\,\mbox{GeV}$ (left panel) and $m_A=800\,\mbox{GeV}$ (right panel), as function of the parameters $y_{l,L}$ (see main text). In both plots we have considered type-I 2HDM with $\tan\beta=1$. The yellow region in the left panel is excluded by present LHC searches. In the region at the left of the 1 fb (left panel) and 0.05 fb (right panel) contours, the production cross-section varies in a negligible way with $y_{l,L}$ and basically coincides with the prediction of the 2HDM without VLLs. The blue region corresponds to theoretically inconsistent, because of RGE effects, values of the Yukawa parameters.}
\label{fig:sigmaRGE}
\end{figure}

\noindent
As already pointed the requirements of a reliable behaviour of the theory under RG evolution affect mostly possible predictions of LHC signals. As it will be reviewed in greater detail in the next subsections, one of the most characteristic signatures induced by the VLLs are enhanced diphoton production rates from decays of resonantly produced $H/A$ states. This happens because their effective couplings with photons are increased by triangle loops of electrically charged VLLs such that, once their masses are fixed, the corresponding rate depends on the size of the Yukawa couplings. The constraints from RGE can be used to put an upper limit on the size of the Yukawa couplings which imply, in turn, an upper limit on the diphoton production cross-sections which are expected to be observed.

\noindent
As illustration we have thus reported in fig.~(\ref{fig:sigmaRGE}) the isocontours of $\sigma( pp \rightarrow A) Br(A \rightarrow \gamma \gamma)$ as function of $y_l=y_h^{E_L}$ and $y_L=y_H^{E_L}=-y_H^{E_R}=-y_H^{N_L}=y_H^{N_R}$ (see below for clarification), for two values of $m_A$, namely 500 and 800 GeV. As further assumption we have set $m_{E_1}=m_A/2$ in order to maximize the effective coupling between $A$ and the photons~\footnote{In the computation we have considered only a ``perturbative'' enhancement. A further enhancement can be achieved through non perturbative effects~\cite{Bharucha:2016jyr}, at the price of a rather strong fine tuning of $|m_A/2-m_{E_1}|$. We won't consider this case in the present work.}.   

\noindent
As it is clear, in order to obtain sensitive deviations from the prediction of a 2HDM without VLLs, which is approximately 1 fb and 0.05 fb for the two examples considered, rather high values of the new Yukawas are needed~\footnote{This requirement can be partially relaxed by introducing more than one family of VLL.}, which would induce too large radiative corrections to the quartic couplings of the scalar potential. In theoretically consistent realizations, the VLLs have negligible effects on the diphoton production cross-section.

\begin{figure}[htb]
\begin{center}
\includegraphics[width=6.5 cm]{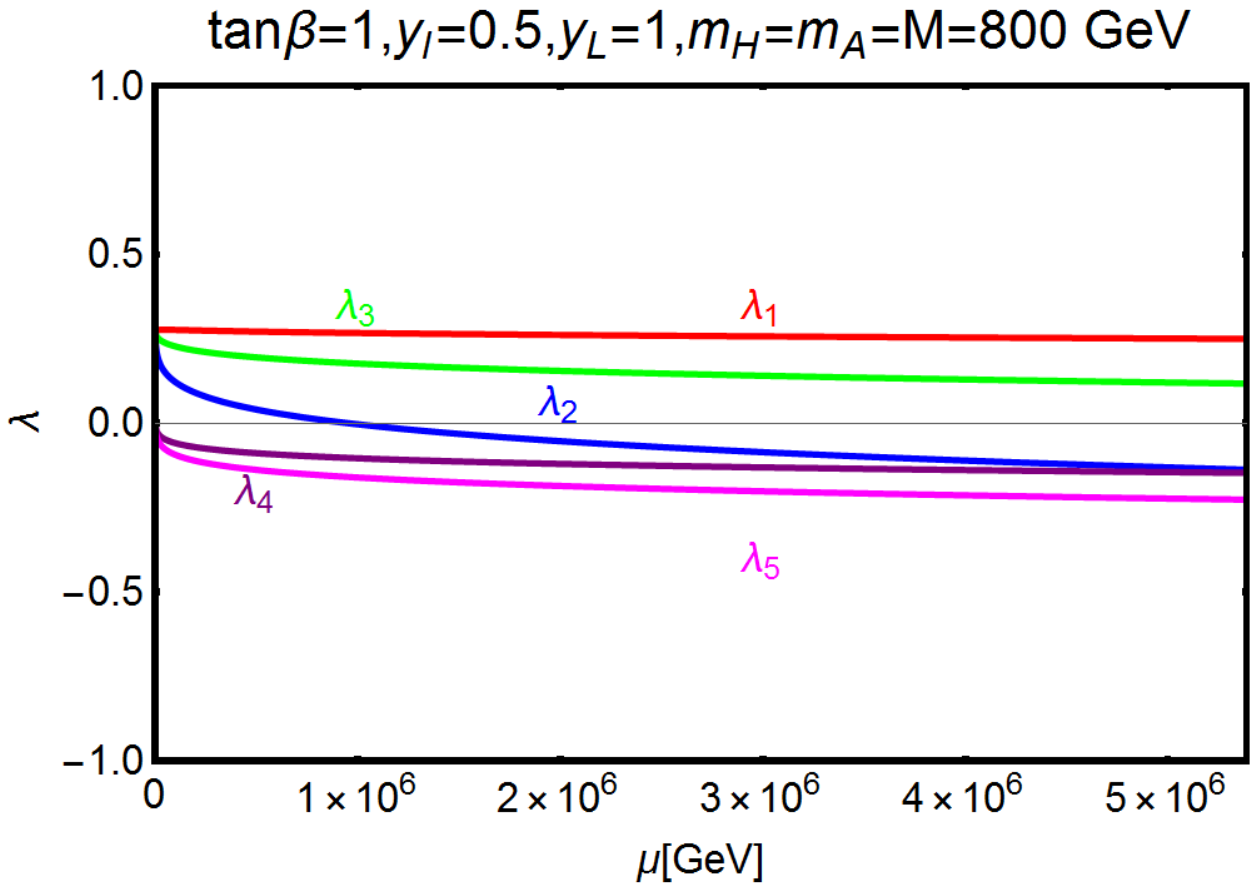}
\includegraphics[width=6.7 cm]{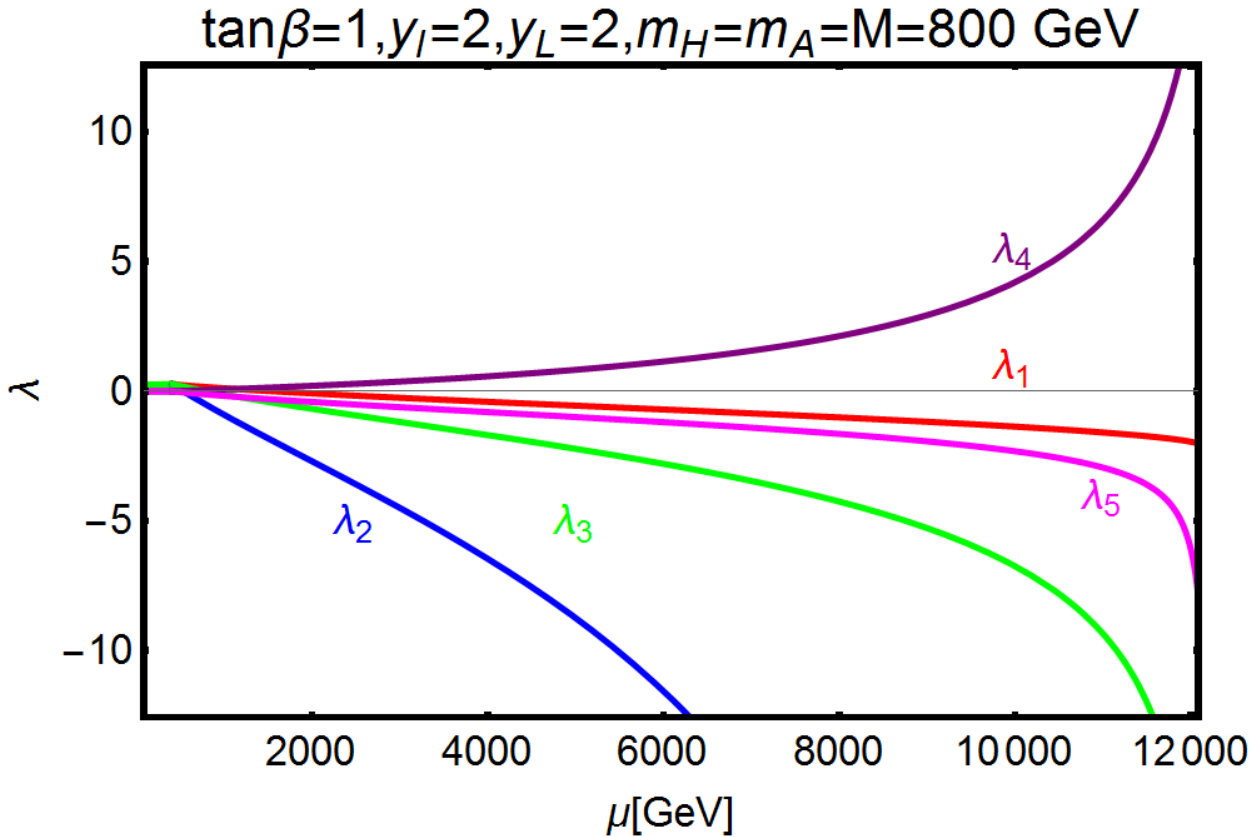}
\end{center}
\caption{Two examples of resolution of the RGE equations. The corresponding assignations of the relevant model parameters are reported on top of the panels. In the left panel the initial values of the Yukawa couplings are sufficiently small such that the conditions~(\ref{eq:up1})-(\ref{eq:up2}) are satisfied up to energy scales of the order of $10^6\,\mbox{GeV}$. In the right panel the assignation of the Yukawas causes, instead, the couplings $\lambda_{1,2}$ to become negative already at the energy threshold of the VLLs.}
\label{fig:RGEexamples}
\end{figure}

\noindent
We have checked the validity of the criteria $|\beta_{\lambda_{1,2}}/\lambda_{1,2}|\leq 1$ by explicitly solving the RGE for some benchmark models. We have reported two sample solutions in fig.~(\ref{fig:RGEexamples}). Here we have considered the same assignation of the model parameters as in the right panel of fig.~(\ref{fig:sigmaRGE}), and chosen two assignations of the input values of the Yukawa parameters $y_{l,L}$. In the left panel we have considered the set $(y_l,y_L)=(0.5,1)$, lying in the white region of the right panel of fig.~(\ref{fig:sigmaRGE}). As evident, the couplings $\lambda_{1,2}$ remain positive up to an energy scale $\mu$ of the order of $10^6\,\mbox{GeV}$, high enough so that the model point is viable at least as a phenomenological description.\footnote{We have explicitly checked the other conditions~(\ref{eq:up1})-(\ref{eq:up2}) as a function of the energy and found that these are violated at a slightly lower scale of $5\times 10^{5}\,\mbox{GeV}$. This difference is acceptable as it does not affect the validity of our results: our goal is not to quantitatively determine the scale at which the theory should be completed, but just to set a qualitative criteria that applies to the theory at low energy.} On the contrary, by considering a parameter assignation lying in the blue region, RGEs drive the couplings $\lambda_{1,2}$ to negative values already at the energy threshold of the charged VLLs, of the order of 400 GeV for the case considered.

\subsection{DM Phenomenology}

\noindent
The coupling of the DM to an additional Higgs doublet has a two-fold impact on dark matter phenomenology. First of all, the extra neutral Higgs states constitute additional s-channel mediators for DM annihilations and, only for the case of $H$, t-channel mediators for scattering processes relevant for Direct Detection. In addition, in the high DM mass regime, they may represent new final states for DM annihilation processes.

\noindent
The coupling of the DM with the non SM-like Higgs states can be expressed, in the mass basis, in terms of the Yukawa couplings $y_{h,H}^X$ and of the mixing angles $\theta_X^{L,R}$: 
\begin{align}
\label{fig:DM_high_H}
& y_{H N_1 N_1}= \frac{\cos \theta_N^L \sin \theta_N^R y_H^{N_L}+\cos \theta_N^R \sin \theta_N^L y_H^{N_R}}{\sqrt{2}}, \nonumber\\
& y_{A N_1 N_1}= i\frac{\cos \theta_N^L \sin \theta_N^R y_H^{N_L}-\cos \theta_N^R \sin \theta_N^L y_H^{N_R}}{\sqrt{2}}, \nonumber\\
& y_{H^+ N_1 E_1}=\cos \theta_N^L \sin \theta_E^R y_H^{E_L}+\sin \theta_N^L \cos \theta_E^R y_H^{E_R}-\cos \theta_N^R \sin \theta_E^L y_H^{N_R}-\cos \theta_N^L \sin \theta_E^R y_H^{N_L}.
\end{align}

\noindent
The analysis of the DM phenomenology is structured in an analogous way as the one performed in the previous section. We will compute the DM annihilation cross-section and verify for which assignations of the parameters of the model the thermally favored value, $\sim 3 \times 10^{-26}$, is achieved without conflicting with bounds from DM Direct Detection. Given the dependence of the coupling between the DM and the neutral Higgs states on the mixing angles $\theta_N^{L,R}$ the DM scattering cross-section is still dominated by the $Z$ exchange processes so that the new couplings from eq.~(\ref{fig:DM_high_H}) mostly impact the determination of the DM relic density.

\noindent
For what regards the DM relic density, we distinguish two cases:
\begin{itemize}
\item $m_{N_1} \leq m_{X}/2,X=A,H,H^{\pm}$. In this case the situation is very similar to the case of SM+VLLs. The most relevant DM annihilation channels are again into fermion and gauge boson pairs. Reminding that, in the alignment limit, there is no tree-level coupling between the $H,A$ states and the $W,Z$ bosons the only annihilation processes sensitively influenced by the presence of the additional Higgs bosons are the ones into SM fermions. In particular, s-channel exchange of the CP-odd Higgs gives rise to a new s-wave contribution so that the DM annihilation cross-section can be schematized as:
\begin{align}
& \langle \sigma v \rangle_{ff}=\frac{m_{N_1}^2}{8 \pi} \frac{m_t^2}{v^2}|\xi_{A}^t|^2 \frac{1}{((4 m_{N_1}^2-m_A^2)^2+m_A^2 \Gamma_A^2)}|y_{A N_1 N_1}|^2 \nonumber\\
& + \frac{g^2 m_{N_1}^2}{\pi ((4 m_{N_1}^2-m_Z^2)^2+m_Z^2 \Gamma_Z^2)} \left[\sum n_c^f (|V_f|^2+|A_f|^2) |y_{V,Z N_1 N_1}|^2\right. \nonumber\\
&\left. +\frac{3 m_t^2}{2 m_{N_1}^2}(|V_t|^2+|A_t|^2)|y_{A,Z N_1 N_1}|^2\right].
\end{align} 
As evident, the annihilation cross-section depends, through the factor $\xi$, on $\tan\beta$ and, in turn, on the realization of the couplings of the two Higgs doublets to SM fermions. Given the dependence on the mass of the final state fermions, $A$-exchange diagrams give a sizable contribution mostly to the $\bar t t$ final state, when kinematically open (an exception being a type II/flipped 2HDM for $\tan\beta \gtrsim 45$, when a sizable contribution comes also from $\bar b b$).

\begin{figure}[htb]
\begin{center}
\subfloat{\includegraphics[width=6.5 cm]{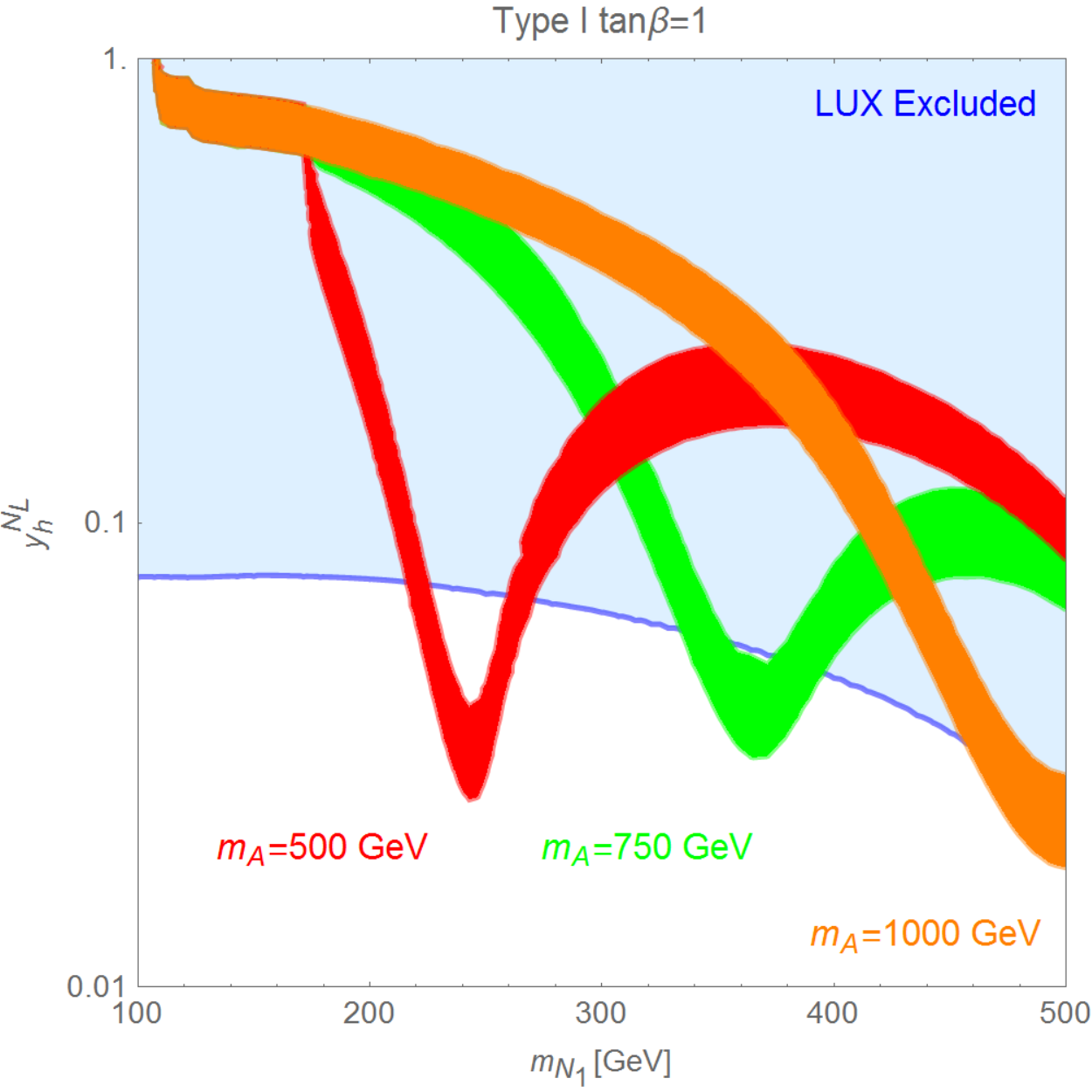}}
\subfloat{\includegraphics[width=6.5 cm]{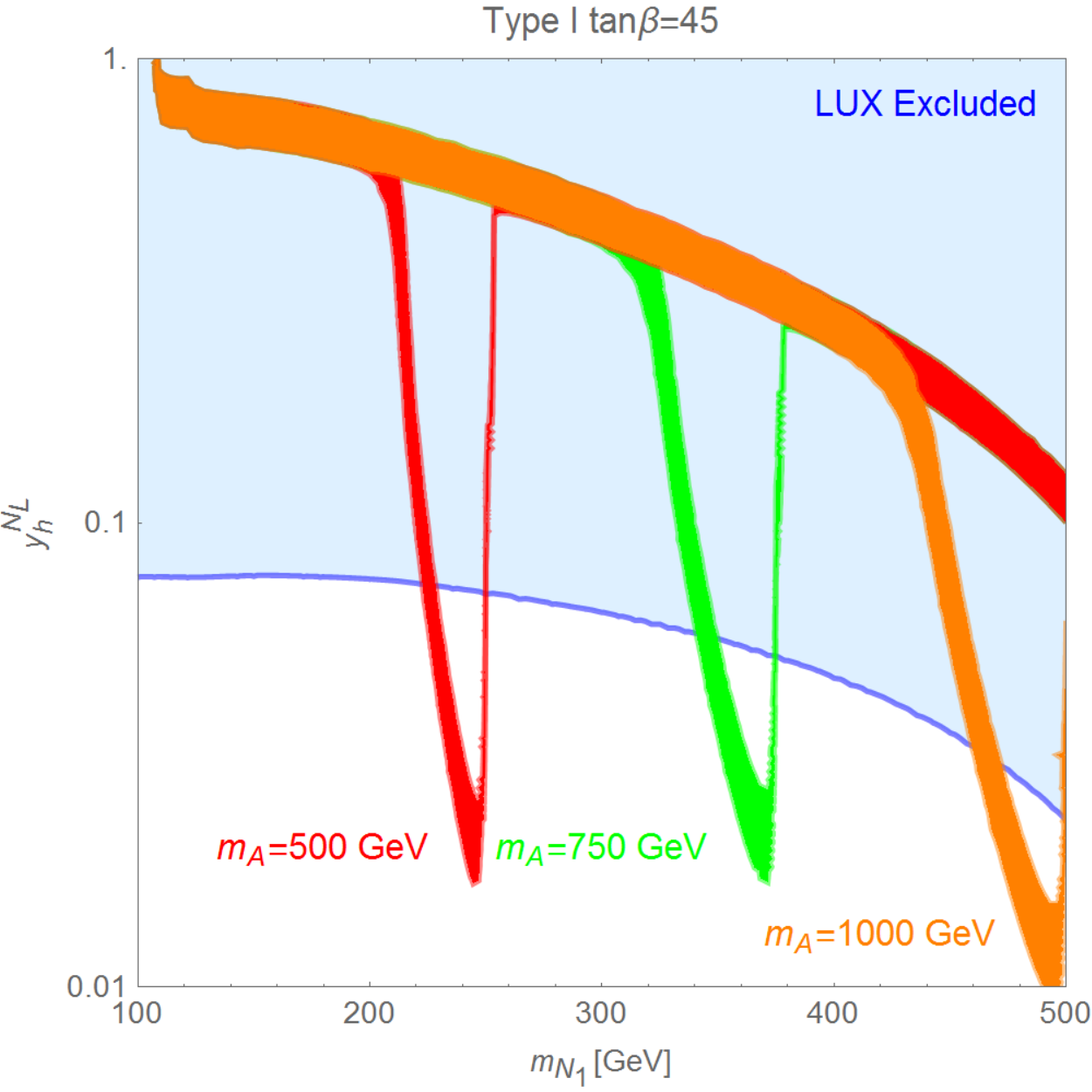}}
\end{center}
\caption{Isocontours of the correct DM relic density in the bidimensional plane $(m_{N_1},y_h^{N_L})$ for two values of $\tan\beta$, (left) 1 and (right) 45, and for the following assignations of the other parameters of the fermion sector: $y_h^{N_R}=y_h^{E_R}=0$, $y_h^{E_L}=0.5$, $y_H^{N_L}=-y_H^{N_R}=-y_H^{E_L}=y_H^{E_R}=1$. We have finally set $M=m_H=m_A=m_{H^{\pm}}$ and considered the three values of 500 GeV, 750 GeV and 1 TeV.}
\label{fig:lightDM}
\end{figure}

\noindent
As already pointed out, the strong DD limits, mostly originating from t-channel $Z$ exchange, impose that the DM is essentially a pure $SU(2)$ singlet with, as well, a tiny hypercharged component. This implies also a suppression of the couplings of the DM to the neutral Higgs states, such that the DM is typically overproduced in the parameter regions compatible with DD constraints. It is nevertheless possible to achieve the correct relic density by profiting of the resonant enhancement of the DM annihilation cross-section when the condition $m_{N_1}\simeq\frac{m_{H,A}}{2}$ is met. Notice that in this case the DM annihilation cross-section is also sensitive to the total width of the $H/A$ state and thus sensitive to the value of $\tan\beta$. An illustration of the DM constraints in the $m_{N_1} \leq \frac{m_{A,H}}{2}$ regime is provided in fig.~(\ref{fig:lightDM}). Here we have compared, for two values of $\tan\beta$ (for definiteness we have considered type-I 2HDM), the isocontours of the correct DM relic density, for three assignations of $m_A=m_H=m_{H^{\pm}}$, and the DD exclusion limit, as set by LUX. As already anticipated the only viable regions are the ones corresponding to the s-channel poles. We also notice that the shapes of the relic density contours are influenced by the large (narrow) widths of the resonances occurring for small (high) $\tan\beta$.   
 
\item $m_{N_1} > m_{X}/2,X=A,H,H^{\pm}$: in this case the situation is very different with respect to the case of the SM Higgs sector. Indeed, as the DM mass increases, new annihilation channels become progressively open. We have, first of all, when $m_{N_1} > m_{X}/2,X=A,H,H^{\pm}$, the opening of annihilation channels of the type $VX$ where $V=Z,W^{\pm},X= A,H,H^{\pm}$. By further increasing the DM mass, annihilation channels into pairs of Higgs states are finally reached. Among these new channels the most efficient turn out to be ones into $W^{\pm} H^{\mp}$ and into $H^{\pm} H^{\mp}$. Indeed this processes can occur through t-channel exchange of the lightest charged state $E_1$ and the corresponding rates depend on the coupling $y_{H^+ N_1 E_1}$ which depends on parameter not involved in direct detection processes and might be of sizable magnitude even for a SM singlet DM, provided that the charged state $E_1$ has a sizable $SU(2)$ component. The potentially rich phenomenology offered by the annihilations into Higgs-Gauge bosons and Higgs boson pairs is the reason why we have not strictly imposed a custodial symmetry in the scalar sector since it would have imposed a too rigid structure to the mass spectrum.  
\end{itemize}

\noindent In order to explore the multi-dimensional parameter space we have then employed a scan of the following parameters:
\begin{align}
& y_h^{N_{L,R}} \in \left[10^{-3},1\right], \nonumber\\
& y_h^{E_{L}} \in \left[5 \times 10^{-3},3\right], \nonumber\\
& M_N \in \left[100\,\mbox{GeV}, 1\,\mbox{TeV}\right], \nonumber\\
& M_E=M_L \in \left[300\,\mbox{GeV}, 1\,\mbox{TeV}\right], \nonumber\\
& \tan\beta \in \left[1,50\right],\nonumber\\
& m_A \in \left[250\,\mbox{GeV},1\,\mbox{TeV}\right],\nonumber\\
& m_H \in \left[m_h, 1.5\,\mbox{TeV}\right],\nonumber\\
& m_{H^{\pm}} \in \left[m_W,1.5\,\mbox{TeV}\right], \nonumber\\
& |M| \in \left[0, 1.5 \,\mbox{TeV}\right],
\end{align}
and required that the model points pass the constraints from EWPT, from perturbativity and unitarity of the scalar quartic couplings, eq.~(\ref{eq:up1})-(\ref{eq:up2}), and from satisfying the requirement of stability under RGEs, $|\beta_{\lambda_{1,2}}/\lambda_{1,2}| <1$. We have finally required that the correct DM relic density is achieved. Similarly to the case discussed in the previous section, we have disregarded the possibility of coannihilations between the DM and other VLLs by further imposing a minimal mass difference between these states. We have repeated this scan for the different 2HDM realizations reported in tab.~(\ref{table:2hdm_type}). Although the DM results are mostly insensitive to the type of couplings of the Higgs states with SM fermions the prospects for LHC searches, discussed in the next subsection, will be different in the various cases.

\begin{figure}[htb]
\begin{center}
\includegraphics[width=8.5 cm]{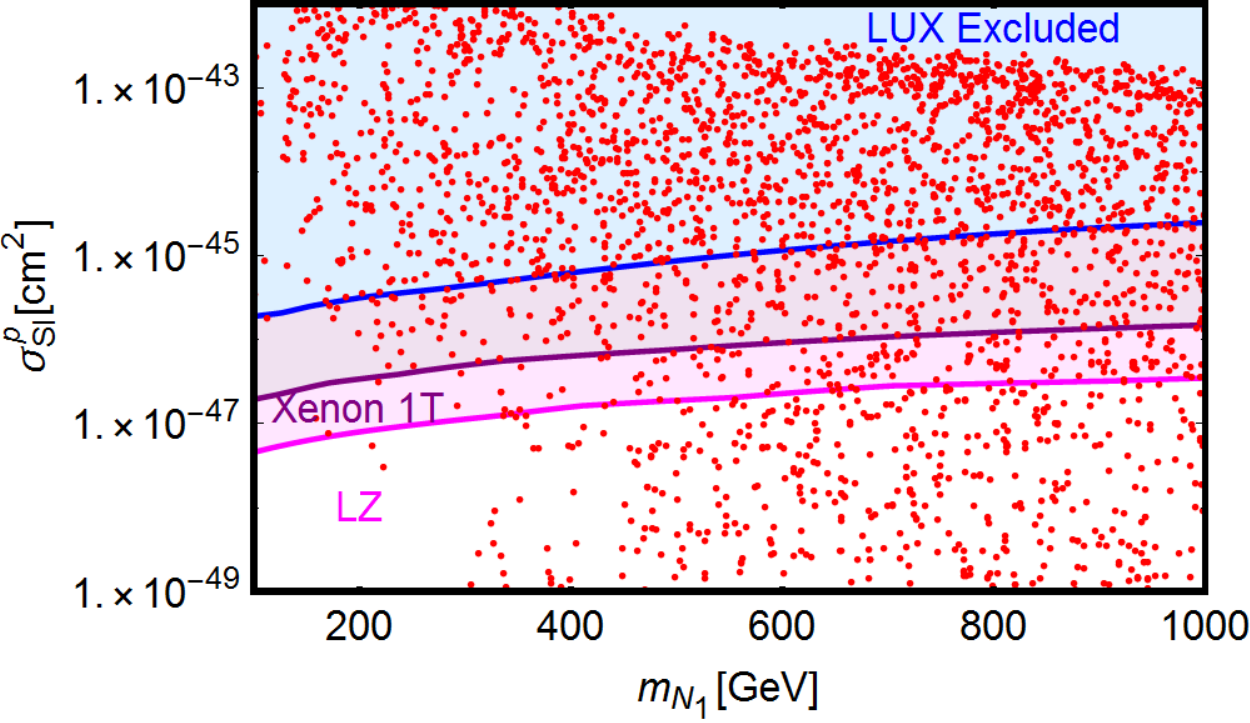}
\end{center}
\caption{Model points satisfying the correct DM relic density and passing EWPT, perturbativity and unitarity constraints, in the bidimensional plane $(m_{N_1},\sigma_{SI})$. The blue region is excluded by current limits by LUX while the Purple and Magenta regions represent the reach of Xenon1T and LZ.}
\label{fig:DMtot}
\end{figure}

\noindent
The results of our analysis have been again reported, in fig.~(\ref{fig:DMtot}), in the bidimensional plane $(m_{N_1},\sigma_{\rm SI})$. Similarly to the case of the single Higgs doublet scenario, many points, especially at lower values of the DM mass, are excluded by LUX. Viable model configurations nevertheless exist, already for DM masses of the order of 150 GeV. We notice in particular the presence of points lying beyond the reach of even next generation 1 Tone facilities like XENON1T and LZ. This is because, for these configurations, the relic density is achieved through the annihilations into $H^{\pm}H^{\mp}$ and $H^{\pm}W^{\mp}$ final states, relying on the couplings $y_{h,H}^{E_{L,R}}$, so that very small values of the neutral Yukawa couplings can be taken (as pointed above, in presence of a single family of VLLs, large deviations from the custodial limit are allowed provided suitable assignments of the masses of the Higgs states.)

\subsection{Impact on LHC}

\noindent
In this section we will discuss the impact on LHC phenomenology of the scenario under investigation. In the subsections below we will provide an overview of the possible relevant processes, which currently are (and will be probed in the near future) by the LHC. These are distinguished in three categories: production of Higgs states and decay into SM fermions; production of the Higgs states and decay into gauge bosons, especially photons; direct production of VLLs. VLLs are directly involved only in the last two categories of collider signals; it is nevertheless important to consider as well limits/prospects from the first category of processes since they put constraints on the masses of Higgs states and on $\tan\beta$ which can, in turn, reduce the viable parameter space for DM.

\noindent
Among this rather broad variety of signals we will dedicate particular attention to the diphoton production. It arises from the resonant production, and subsequent decay into photon pairs, of the neutral Higgs states. The VLL couplings entering in this process are the Yukawa couplings $y_{h,H}^{E_{L,R}}$. These couplings control the annihilation cross-sections into $W^{\pm}H^{\mp}$ and $H^{\pm} H^{\mp}$ final states, which mostly account for the DM relic density in the high mass regime; furthermore, they are influenced, through the $S/T$ parameters, by the values of the neutral couplings $y_{h,H}^{N_{L,R}}$, which are in turn strongly constrained by DM phenomenology.

\noindent
As further simplification we will consider the CP-even Higgs state $A$ as the only candidate for a diphoton resonance. As it will be explicitly shown in the following, this condition can be achieved by imposing a specific relation between the VLF Yukawa couplings, so as to minimize the impact of VLLs on the effective couplings between the CP-even state $H$ and photons and, at the same time, maximize their impact on the effective $A\gamma\gamma$ coupling. This relation will allow to reduce the number of free parameters. This choice is also motivated by the fact that the production cross-section  $pp \rightarrow A$ of the CP-odd state is, at parity of masses, bigger than the corresponding one of the CP-even state $H$. For the specific case of the diphoton production, as already pointed out, a further enhancement is achieved by a specific choice of the masses of the charged VLLs. As a consequence, focusing on the CP-odd Higgs $A$ allows to obtain conservative limits which can be straightforwardly extended to the CP-even $H$.

\noindent
Despite these simplifications, there is still a broad variety of factors which influence the collider phenomenology of a diphoton resonance. We thus summarize below the most relevant cases, basically distinguished by the value of $\tan\beta$: 

\begin{itemize}
\item {\bf Low} $ \tan  \beta$, i.e $\tan \beta=1-7$: The neutral Higgs states are mostly produced through gluon fusion. Irrespective of the type of couplings with the SM fermions (see table~\ref{table:2hdm_type}), the top coupling to the heavy scalars is the dominant among the ones with SM fermions. This last coupling determines almost entirely the production cross sections of the processes $pp \rightarrow A/H$. The $H/A$ resonances would then dominantly decay into $\bar t t$, or into a lighter neutral scalar (whether kinematically allowed) and a gauge boson,\footnote{This possibility is contrived because the very strong $HAZ$ coupling would easily lead to very high decay widths, which would make difficult the observation of resonances.} except for the case of sizable branching fractions of decay into charged and neutral VLLs (an important branching fraction into the DM would be nevertheless in strong tension with constraints from DM searches). In particular, for $\tan\beta=1$, one can have very large, $\Gamma/M \sim 5-10 \%$, decay width, given essentially by decays into $\bar t t$. The observation of $t \bar t$ resonances would be an interesting complementary signature of an eventual diphoton resonance. Searches of this kind of signals have been already performed at LHC Run I~\cite{Khachatryan:2015sma,Aad:2015fna}. The gluon-gluon fusion (ggF) mechanism can provide production cross-sections close to the experimental sensitivity only for $\tan\beta \simeq 1$, while for increasing values of $\tan\beta$ it gets rapidly suppressed.
 
\item  {\bf Moderate} $\tan  \beta$, i.e. $\tan \beta=10-20$: While gluon fusion is still the most relevant production process, in a 2HDM with enhanced $\xi^d_{H,A}$ (type II and lepton-specific), a sizable contribution arises also from $b \bar b$ fusion. Regardless of the type of the 2HDM, the couplings between neutral resonances and SM fermions are suppressed, with respect to the previous scenario, so that they feature rather narrow width, unless sizable contributions arise from decays into VLLs (for $\tan \beta \gtrsim 5$ unitarity and perturbativity constraints favor a degenerate Higgs spectrum.). Large cross sections for the process $pp \rightarrow A/H \rightarrow \tau \tau$ are expected in a 2HDM with enhanced $\xi^l_{H,A}$, i.e. type II and flipped. Corresponding LHC searches~\cite{Aad:2014vgg,Khachatryan:2014wca} give already strong limits, such that values of $\tan\beta$ above 10 are already excluded for $m_{A,H} < 500\,\mbox{GeV}$. 

\item  {\bf High} $ \tan  \beta$, i.e. $\tan \beta \simeq 50$: This regime occurs only for the type-I and flipped 2HDM since the other cases are essentially ruled out, for masses of the neutral Higgses below approximately 1 TeV, by the limits from $pp \rightarrow A/H \rightarrow \tau \bar \tau$. Two rather different scenarios correspond to these two types of 2HDM. In the flipped model the $A/H$ Higgs have enhanced couplings with $b$-quarks, implying $b \bar b$-fusion as dominant production process and, possibly, a large decay width dominated by the $b \bar b$ final state. In the case of the type-I 2HDM the neutral Higgses are ``fermiophobic'', since all their couplings to the SM fermions are suppressed by a factor $1/\tan\beta$. Unless the decays into VLLs are relevant, we have very narrow widths, even $\Gamma_{H,A}/m_{H,A} \sim 10^{-2}$, and a strong enhancement of the decay branching fraction into photons.
\end{itemize}

\noindent
In the following subsections we will provide an overview, for the scenarios depicted above, of the possible relevant LHC signals and the corresponding constraints/prospects of detection. We have indeed identified some relevant subsets among the parameter points providing the correct DM relic density and in agreement with theoretical constraints. We have first of all considered a set of points in the low, namely $1-5$, $\tan\beta$ regime (although we will mostly refer to type-I 2HDM, the various 2HDM realizations do not substantially differ in this regime, as already pointed out). To these we have added three subsets, characterized by $10 \leq \tan\beta \leq 40$, for, respectively, type-I, type-II and lepton-specific couplings of the 2 Higgs doublets with the SM fermions. Two subsets at $\tan\beta=50$, corresponding to type-I and flipped realizations, have been finally included. 

\noindent
For our study we have adopted the cross-sections provided by the LHC Higgs Cross Section Working Group~\cite{deFlorian:2016spz}, which have been produced with \texttt{SusHi}~1.4.1~\cite{Harlander:2012pb}. More specifically, for the 2HDM types with enhanced bottom quark couplings to heavy scalars (type-II and flipped), we have taken the $gg/\bar bb$ fusion cross sections calculated for the $h$MSSM~\cite{Djouadi:2013uqa,Djouadi:2015jea}. For the remaining two realizations, namely type-I and lepton-specific 2HDMs, regardless of the value of $\tan\beta$, the only important production mechanism is $gg$ fusion, since $\bar bb$ fusion is suppressed not only by the lower bottom quark luminosity, but also by the $\bar bbA/H$ couplings, which scale as $1/\tan\beta$. Therefore, as both top and bottom quark couplings to the heavy scalars are proportional to $1/\tan\beta$ for type-I and lepton specific 2HDMs, it follows that the effective $ggA/H$ couplings have a similar behaviour. Consequently, for these two realizations, we evaluated the $gg$ fusion cross sections by simply taking the $h$MSSM $gg$F cross section for $\tan\beta = 1$ and rescaling it by $1/\tan^2\beta$.

\subsubsection{$A/H \rightarrow \bar f f$}

\noindent
We will start our analysis by considering the production processes $pp \rightarrow \bar f f$.

\noindent 
Their phenomenology is virtually identical to the pure 2HDM case. Indeed, being singlets under $SU(3)$, the VLF do not modify the gluon fusion production vertex; furthermore, as better clarified in the next subsections, their effect on the decay branching fraction of the Higgs states is mostly negligible.

\noindent
Since the branching fractions of the Higgses decaying into fermions depend on the masses of the final state fermions themselves, sizable signals can be achieved only for $\bar t t$, $\tau \tau$ and $\bar b b$ final states. The observation of the latter is substantially precluded by huge SM backgrounds so only $\bar t t$ and $\tau \tau$ feature observational prospects. Tau pair searches can probe type-II 2HDMs at moderate-to-high $t_{\beta}\gtrsim 5$, depending on the value of $m_A$, since in this case we have an enhancement of the $\tau$ Yukwawa coupling to $A$, $\xi^{\tau}_A = t_{\beta}$. In a complementary manner, $t\bar{t}$ searches provide a discovery avenue for small values of $t_{\beta}$, typically $\lesssim 3$~\cite{Carena:2016npr}, for any type of 2HDM. However, looking for heavy scalars decaying into top quark pairs is challenging from the experimental point of view, since the interference between the signal and the SM background can give rise to non-trivial dip-peak structures in the $\bar{t}t$ invariant mass spectrum, which get smeared after binning, thus reducing the visibility of a potential ``bump''~\cite{Carena:2016npr,Djouadi:2016ack}. We also mention that the search 
for scalar resonances lighter than $500$~GeV decaying to $\bar{t}t$ pairs is not possible, as the $t$ and $\bar t$ quark are not boosted enough, the selection cuts thus being inefficient. 

\begin{figure}[htb]
\begin{center}
\includegraphics[width=8 cm]{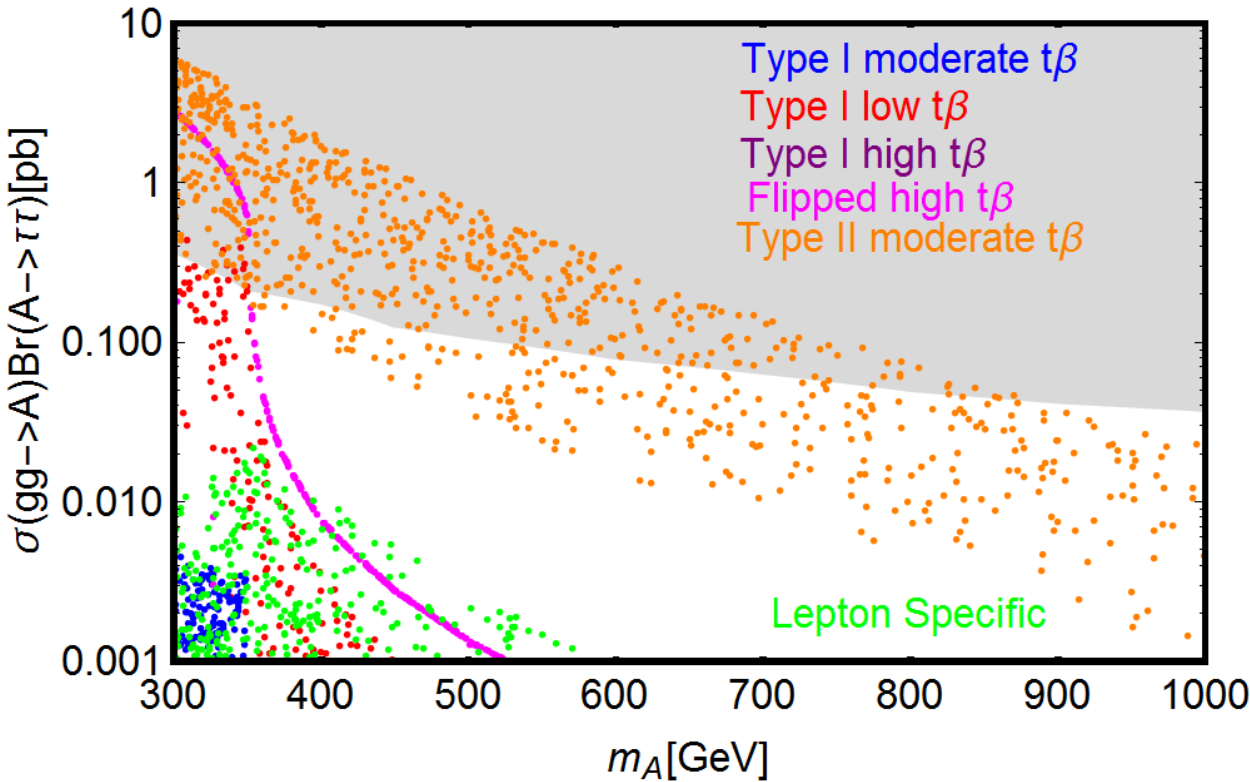}
\end{center}
\caption{Production cross-section for the process $pp \rightarrow \bar \tau \tau$ for the set of models with viable relic density. The colors distinguish the type of 2HDM realizations. The gray region is excluded by current limits~\cite{Aad:2014vgg,Khachatryan:2014wca}.}
\label{fig:tautauI}
\end{figure}

\noindent
We have reported in fig.~\ref{fig:tautauI} the $\tau \tau$ production cross-section for the model points passing theoretical and DM constraints, distinguishing, with different colors, the various 2HDM scenarios depicted above. As already stated, current LHC constraints are mostly efficient in the 2HDM-II. They can nevertheless also exclude low values of $m_A$  for other 2HDM realizations.

\begin{figure}[htb]
\begin{center}
\includegraphics[width=7.0 cm]{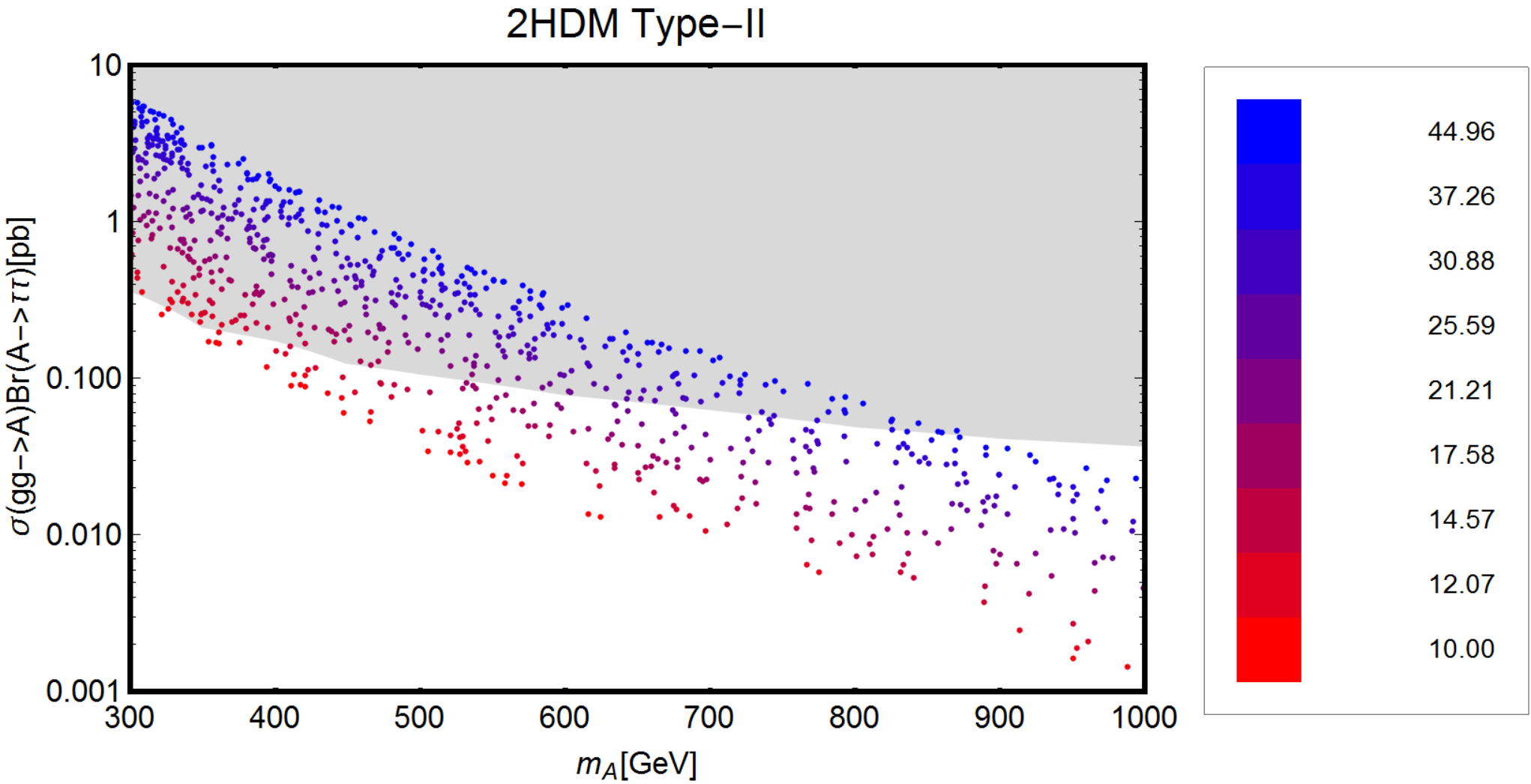}
\hspace{5 mm}
\includegraphics[width=7.0 cm]{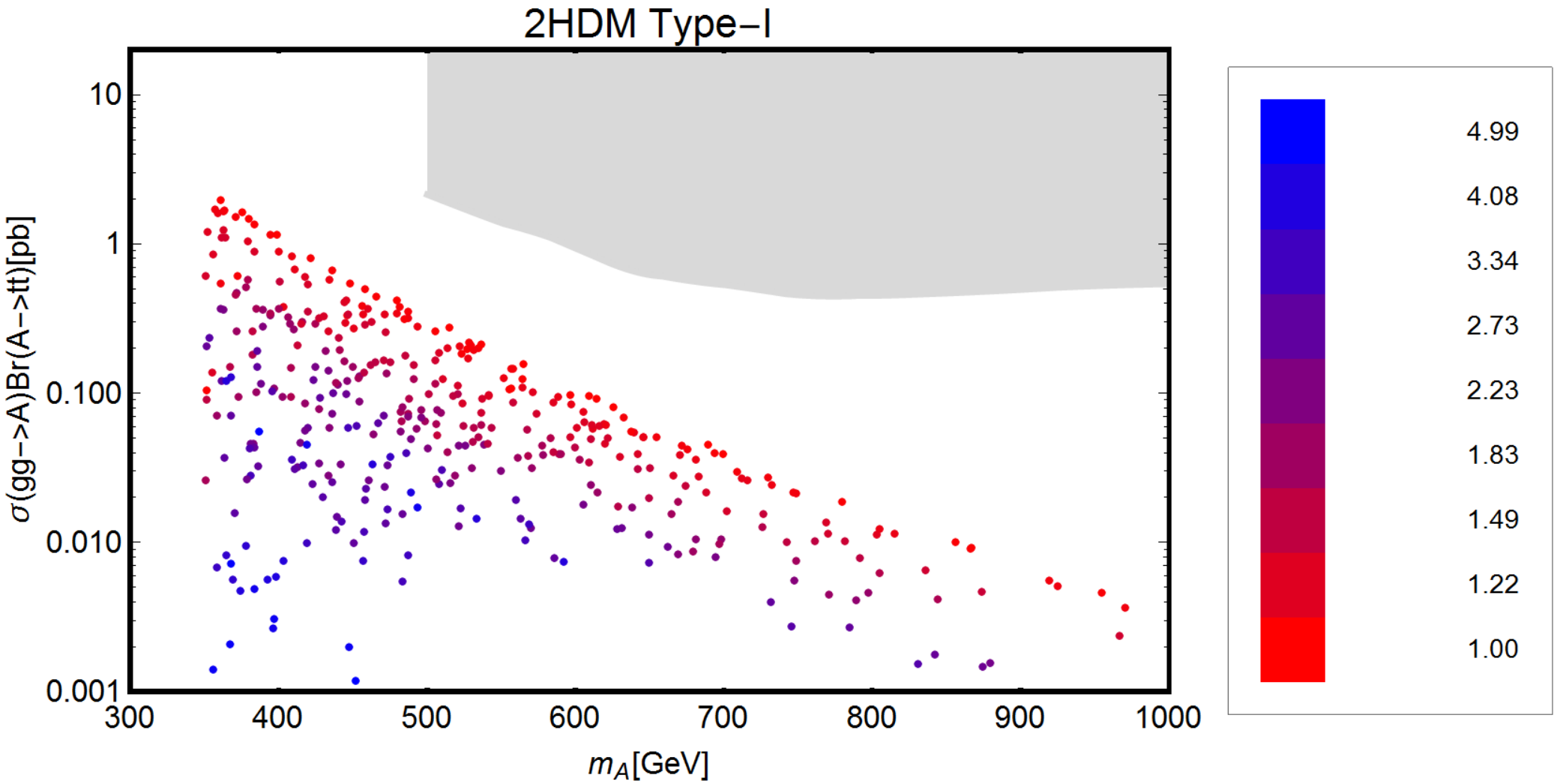}
\end{center}
\caption{Left panel: $pp \rightarrow \bar \tau \tau$ cross-section for type-II 2HDM. Left panel: $pp \rightarrow \bar t t$ for 2HDM type-I realizations in the low $\tan\beta$ regime. In both plots the points follow a color code according to the value of $\tan\beta$. The gray regions are already experimentally excluded.}
\label{fig:Aff}
\end{figure}

\noindent
We have then focused, on the left panel of fig.~(\ref{fig:Aff}), on the 2HDM-II case, highlighting the dependence of the collider limits on the value of $\tan\beta$. As evident, value above 20 are excluded for $m_A$ up to 1 TeV. A similar exercise has been performed on the right panel of fig.~(\ref{fig:Aff}) for the case of the $pp \rightarrow A \rightarrow \bar t t$ process, in the scenario of very low $\tan\beta$. As evident, all the points lie below current experimental sensitivity. Only the points with $\tan\beta \sim 1$ lie close enough to the experimental sensitivity in order to be probed by in the next future.

\subsubsection{Diphoton Signal}

\noindent
In this subsection we will investigate in more detail the prospects for observing a diphoton signal. The corresponding cross-section can be schematically written as:
\begin{equation}
\sigma (pp \rightarrow \Phi \rightarrow \gamma \gamma)=\sigma (pp \rightarrow \Phi) Br(\Phi \rightarrow \gamma \gamma), \,\,\,\,\Phi=H,A,
\end{equation}
with
\begin{equation}
Br(\Phi \rightarrow \gamma \gamma) \propto |\mathcal{A}_{\rm SM}^\Phi+\mathcal{A}_{H^{\pm}}^\Phi+\mathcal{A}_{\rm VLL}^\Phi|^2,
\end{equation}
where $\mathcal{A}_{\rm SM}^\Phi$, $\mathcal{A}_{H^{\pm}}^\Phi$ and $\mathcal{A}_{\rm VLL}^\Phi$ represent, respectively, the loop induced amplitudes by SM fermions, charged Higgs (only present for the CP-even state $H$) and VLLs.

\noindent
The contribution associated to the VLLs can be written as:
\beq
\mathcal{A}_{\rm VLL}^{\Phi} = \sum_{i=1}^2 \frac{v \left(\mathcal{C}_E^{\Phi}\right)_{ii}}{m_{E_i}} A_{1/2}^{\Phi} (\tau_{E_i}),
\label{eq:generic_ampl}
\eeq 
where we have used the definition:
\beq
\mathcal{C}_E^{\Phi} = U_L^E \cdot \mathcal{Y}_E^{\Phi} \cdot \left( U_R^E \right)^{\dagger}.
\eeq
The Yukawa couplings between the VLLs and the heavy Higgs states are given by
\beq
\mathcal{Y}^H_N = \frac{1}{\sqrt{2}} \left( \begin{array}{cc} 
0 & y_H^{N_L} \\ 
 y_H^{N_R} & 0 
\end{array}\right), \quad 
\mathcal{Y}^H_E = \frac{1}{\sqrt{2}} \left( \begin{array}{cc} 
0 & y_H^{E_L} \\ 
 y_H^{E_R} & 0 
\end{array}\right),
\eeq
for the heavy CP-even scalar $H$ and:
\beq
\mathcal{Y}^A_N = \frac{1}{\sqrt{2}} \left( \begin{array}{cc} 
0 & -y_H^{N_L} \\ 
 y_H^{N_R} & 0 
\end{array}\right), \quad 
\mathcal{Y}^A_E = \frac{1}{\sqrt{2}} \left( \begin{array}{cc} 
0 & y_H^{E_L} \\ 
 -y_H^{E_R} & 0 
\end{array}\right),
\eeq
for the CP-odd scalar $A$.

\noindent
A general analytical expression for eq.~(\ref{eq:generic_ampl}) would be rather involved. We will however consider two simplifying assumptions. First of all, in order to avoid dangerous contribution to the decay branching fraction into photons of the SM-like Higgs we will set, as done before, $y_h^{E_R}=0$. Note that, especially in the case of heavier VLLs, one can relax this assumption, since the $h\to\gamma\gamma$ signal strength is currently measured with only $\sim 10-20\%$ accuracy; nevertheless, for simplicity, we will take $y_h^{E_R}=0$. Furthermore, we will assume $M_E=M_L$, such that the mass matrix for the charged VLLs simplifies to~\footnote{In fact, we checked that such a texture for the mass matrix suppresses, in a similar way as for $h\to\gamma\gamma$, the VLL contributions to $h\to Z\gamma$, thus leaving the latter decay SM-like.}
\beq
\mathcal{M}_E = \left( \begin{array}{cc} 
M_E & v' y_h^{E} \\ 
0 & M_E 
\end{array}\right).
\eeq
Knowing that neither the sign of $M_E$ nor the one of $y_h^{E}$ are physical (both signs can be absorbed via a field redefinition), we will consider only positive values for these parameters. Thus, the eigenmass splitting reads
\beq
m_{E_2} - m_{E_1} = v' y_h^E,
\eeq
with $M_E = \sqrt{m_{E_1} (m_{E_1} + v' y_h^E)}$ fixed in order to give $m_{E_1}$ as the lowest eigenvalue. Under this assumptions the heavy scalar loop amplitudes can be written as: 
\begin{align}
  \mathcal{A}_{\rm VLL}^H  &= \frac{-v'}{2 m_{E_1} + v' y_h^E} \bigg\lbrace  y_H^{E_L} \left[ A_{1/2}^H (\tau_{E_1}) - A_{1/2}^H (\tau_{E_2}) \right] \notag \\ 
				   &+ y_H^{E_R} \left[ \frac{ m_{E_1} + v' y_h^E}{m_{E_1}} A_{1/2}^H (\tau_{E_1}) - \frac{ m_{E_1} }{m_{E_1} + v' y_h^E } A_{1/2}^H (\tau_{E_2}) \right] \bigg\rbrace, \\
  \mathcal{A}_{\rm VLL}^A  &= \frac{-v'}{2 m_{E_1} + v' y_h^E} \bigg\lbrace  y_H^{E_L} \left[ A_{1/2}^A (\tau_{E_1}) - A_{1/2}^A (\tau_{E_2}) \right] \notag \\ 
				   &- y_H^{E_R} \left[ \frac{ m_{E_1} + v' y_h^E}{m_{E_1}} A_{1/2}^A (\tau_{E_1}) - \frac{ m_{E_1} }{m_{E_1} + v' y_h^E } A_{1/2}^A (\tau_{E_2}) \right] \bigg\rbrace.
\end{align}
To improve the detection potential of the heavy scalars decaying into diphotons, one should maximize the value of $\mathcal{A}^A_{\rm VLL}$.  This task is achieved by taking opposite signs for the $ y_H^{E_R}, y_H^{E_L}$ couplings. We can thus reduce the number of free couplings by setting $ y_H^{E_R} = - y_H^{E_L} \equiv y_H^E$. In this setup the $H$ and $A$ loop amplitudes become: 
\begin{align}
\mathcal{A}^H_{\rm VLL} &= \frac{-v'^2 y_h^E y_H^E}{m_{E_1} (2 m_{E_1} + v' y_h^E)} \left[ A_{1/2}^H (\tau_{E_1}) + \frac{m_{E_1}}{m_{E_1} + v' y_h^E} A_{1/2}^H (\tau_{E_2})\right], \\
\mathcal{A}^A_{\rm VLL}  &= \frac{v' y_H^E}{m_{E_1}} \left[ A_{1/2}^A (\tau_{E_1}) - \frac{m_{E_1}}{m_{E_1} + v' y_h^E} A_{1/2}^A (\tau_{E_2})\right].
\label{ampl_simplified}
\end{align}
Note that, in the case where both $E_{1,2}$ mass eigenstates are much heavier than the scalar masses, i.e. $\tau_{E_{1,2}} \to 0$, the CP-even and CP-odd amplitudes differ only through the loop form factor:
\beq
\mathcal{A}^{A/H}_{\rm VLL} \simeq \frac{\pm v'^2 y_h^E y_H^E}{m_{E_1} ( m_{E_1} + v' y_h^E)} A_{1/2}^{A/H} (0). 
\eeq
However, in the case where $A_{1/2}^A (\tau_{E_1})$ dominates over the second term in the brackets from eq.~\eqref{ampl_simplified}, which happens, for example, if $m_{E_1}\simeq m_A/2$ and $m_{E_2} \gg m_{E_1}$, the CP-odd amplitude is indeed maximized: $\mathcal{A}^A_{\rm VLL} \propto \frac{v'}{m_{E_1}}$, whereas $\mathcal{A}^H_{\rm VLL} \propto \frac{v'^2}{m_{E_1}m_{E_2}}$.

\begin{figure}[htb]
\begin{center}
\includegraphics[width=8 cm]{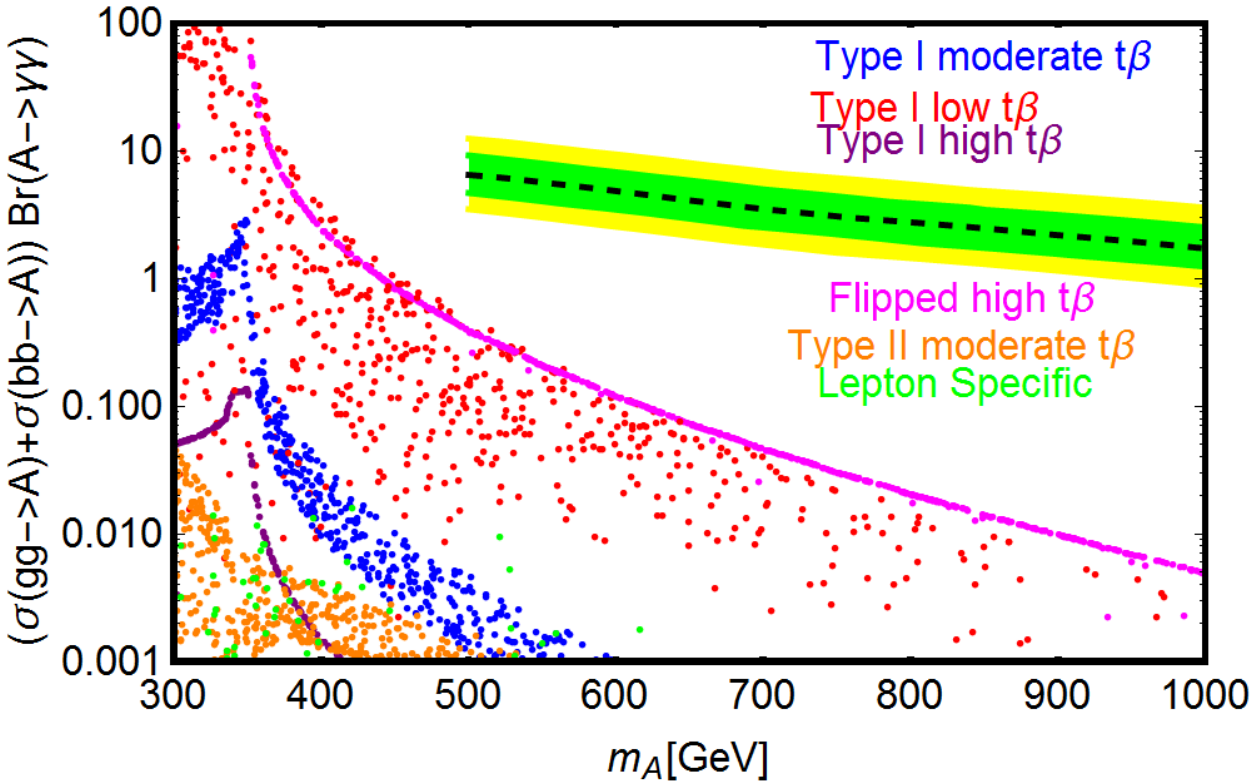}
\end{center}
\caption{Expected diphoton cross-section, as function of $m_A$ for the model points featuring the correct DM relic density and pass constraints from EWPT, perturbativity and unitarity. The red points refer to type-I couplings of the Higgs doublets while the blue ones to the other type of couplings considered in this work.}
\label{fig:pgammagamma}
\end{figure}

\noindent
We have reported, and confronted with the current experimental limits~\cite{CMSICHEP16}, in fig.(\ref{fig:pgammagamma}) the predicted cross-section for $pp \rightarrow A \rightarrow \gamma \gamma$, for the model points providing viable DM candidates. We have distinguished between the different regimes described in the previous subsection, identified by the type of interactions with the fermions and by the value of $\tan\beta$. As evident, the most promising scenarios are the ones corresponding to low $\tan\beta$ and to $\tan\beta \sim 50$ for the flipped 2HDM. These scenarios correspond, indeed, to the configurations which maximize the production vertex of the resonance: as already emphasized, for $\tan \beta \sim 1$ the gluon fusion process is made efficient by the coupling with the top quark, while for $\tan\beta \sim 50$ the production cross-section is enhanced by $b$-fusion. In the other type-I regimes, the cross-section fastly drops with the value of $\tan\beta$.

\noindent
In all the regimes considered the diphoton cross-section lies below the current experimental sensitivity; the deviation from experimental sensitivity fastly reaches several orders of magnitude as the value of $m_A$ increases. A signal in diphoton events would be hardly observable, even in future luminosity upgrades, for $m_A \gtrsim 700\,\mbox{GeV}$. The reason of this outcome mostly lies on the fact that the size of the Yukawa couplings of the charged VLLs are limited from above by the requirement of consistency under RG evolution and, only for $y_h^{E_L}$, by EWPT. As a consequence, no sensitive enhancement of the diphoton production cross-section, with respect to the 2HDM without VLLs, is actually allowed. We notice, in addition, that in order to comply with limits from DM phenomenology, the VLLs should be typically heavier than the diphoton resonance. This translates in a further suppression of the VLL triangle loop contribution.

\subsubsection{Other Loop-Induced Processes}

\noindent
Given their quantum number assignments (and gauge invariance), VLLs also induce, at one loop, decays of $A/H$ into $Z\gamma, ZZ, WW$, which can be probed at the LHC.  

\noindent
Among these processes, the cleanest signal is likely provided by the $Z\gamma$ channel. It is searched for in events with one photon and dijets or dileptons originating from the decay of the $Z$. Although the corresponding production rate is suppressed with respect to diphoton signals, the potential signal is particularly clean (i.e. low background), especially in the case of lepton final states. In the setup under investigation, the $A\to Z\gamma$ decay width, to a very good approximation, reads~\cite{Djouadi:2005gj,Bizot:2015zaa}
\beq
\Gamma(A\to Z\gamma) = \frac{\alpha g^2 m_A^3}{512 \pi^4 v^2 c_W^2}\left( 1-\frac{m_Z}{m_A} \right)^3 \left| \mathcal{A}_t^{AZ\gamma} + \mathcal{A}_b^{AZ\gamma} + \mathcal{A}_{\rm VLL}^{AZ\gamma} \right|^2.
\label{eq:AtoZgamma}
\eeq
The top-loop and bottom-loop amplitudes have simple expressions,
\beq
\mathcal{A}_{t,b}^{AZ\gamma} = N_c Q_{t,b} \, g_V^{t,b} \,  \xi^{t,b}_A A_{1/2}^A(\tau_{t,b}, \lambda_{t,b}),   
\eeq
with $Q$ the electric charge of the SM fermion, $g_V$ its vectorial coupling to the $Z$ boson, and $\xi_A^{t,b}$ defined in Table~\ref{table:2hdm_type}. For the $A_{1/2}^A(\tau_i, \lambda_i)$ loop form factors, we use the same expressions as in Ref.~\cite{Bizot:2015zaa}, with $\tau_i \equiv \frac{m_A^2}{m_i^2}$ and $\lambda_i \equiv \frac{m_Z^2}{m_i^2}$.

\noindent
Concerning the VLL $A\to Z\gamma$ loop amplitude, its general expression, which is again given in the Appendix of Ref.~\cite{Bizot:2015zaa} (denoted as $\tilde{\mathcal{A}}_f^{Z\gamma}$ there), is rather contrived, and will not be displayed here. However, for our particular choice of the charged VLL mass and pseudoscalar Yukawa matrices, it takes the simple form
\beq
\mathcal{A}_{\rm VLL}^{AZ\gamma} = Q_e \, g_V^e \frac{v' y_H^E}{m_{E_1}} \left[ A_{1/2}^A(\tau_{E_1}, \lambda_{E_1}) - \frac{m_{E_1}}{m_{E_1} + v' y_h^E} A_{1/2}^A(\tau_{E_2}, \lambda_{E_2}) \right],
\eeq
with $Q_e = -1$ the electric charge of the VL electron and $g_V^e = -0.25 + s_W^2$ the vectorial coupling to the Z of the SM electron. One can see that, contrary to the general case, the diagrams with off-diagonal $A$ and $Z$ couplings to the VLFs vanish for our choice of parameters. Unfortunately, due to the smallness of $g_V^e \simeq 0.02$, our scenario does not produce a sizeable modification to the $A\to Z\gamma$ decay channel with respect to the case of an ordinary 2HDM. 

\noindent
We also briefly comment on the case of the $WW$ and $ZZ$ decay channels. As the $A\to \gamma\gamma/Z\gamma$ processes,
both $A\to ZZ$ and $A\to WW$ are loop-suppressed ($AWW/AZZ$ vertices are forbidden at tree level by CP-invariance). Moreover, detection of such decays is challenging due to
either (i) suppression by reduced branching ratios (${\rm Br}(Z\to\ell^+\ell^-) \simeq 7\%$, $\ell=e,\mu$) or (ii) final states that are difficult to reconstruct/disentangle from the background  (hadronic decays of  $W,Z$ and leptonic decays of the $W$, $W\to \nu\ell$, which involve missing transverse energy). Therefore, we will not consider these channels as they are not as clean and/or competitive as the ones already discussed.

\subsubsection{Direct Production of VLLs}

\noindent
We conclude our overview of the collider phenomenology of the scenario under investigation by briefly commenting on possible direct searches of the VLLs. VLLs can be produced at LHC through the Drell-Yann processes~\cite{Djouadi:2016eyy} $pp \rightarrow Z^{*}/\gamma^{*} \rightarrow EE$, $pp \rightarrow Z^{*} \rightarrow NN$, and $pp \rightarrow W^{*} \rightarrow NE$. The results of corresponding LHC searches~\cite{Aad:2015cxa,Aad:2015dha} cannot be, nevertheless applied to our case since they rely on the presence of a mixing with SM leptons. Whereas, in our scenario, in order to guarantee the stability of the DM candidate, we have forbidden such a mixing by imposing a $\mathbb{Z}_2$ symmetry under which the VLL sector is odd and the SM is even. On the contrary, a possible collider signal would be represented by the production of $E_1 E_1$ or $N_2 E_1$ and their subsequent decay into DM, which can be tested in 2-3 charged leptons plus missing energy final state events. Direct production of DM, through off-shell Z/h boson or on-shell heavy Higgses, cannot be instead tested, through monojet searches, since constraints from DM Direct Detection imposes, for these states, a negligible branching fraction of decay into DM pairs.

\noindent Another potentially interesting channel would be the production of a charged Higgs and its subsequent decay into $N_1 E_1$, followed by $E_1 \to N_1 W$. However, for most of the points providing the correct DM relic density and, at the same time, passing the DD constraints, we have that $m_{H^{\pm}} < m_{N_1}+m_{E_1}$, so that production can occur only through off-shell charged Higgs. Furthermore, the dominant production modes of $H^{\pm}$ at the LHC, $gg \to t b H^{\pm}$ and $gb \to t H^{\pm}$, are phase-space suppressed by the top quark produced in association and typically have a low cross section. The $s$-channel production of a charged Higgs, $qq' \to H^{\pm}$ is not a valid option neither: even if the charged Higgs would be on-shell, the low Yukawa couplings of the initial state quarks renders such a process unobservable. For a more detailed discussion, we refer the reader to Ref.~\cite{Djouadi:2016eyy}.

\noindent We close the section by a brief comment regarding the possibility of producing long-lived (on a collider timescale) charged particles. Since $N_2$ and $E_2$ both have a sizeable admixture of VL leptons charged under $SU(2)_L$, they almost always decay promptly into $E_1$ plus a $W/Z/h$ boson, the former having a strong $SU(2)_L$ doublet component as well. On the contrary, $N_1$ is a $SU(2)_L$ singlet with a tiny doublet admixture, which greatly suppresses the $E_1 \to N_1 W$ decay rate. Nevertheless, as we imposed a sizeable mass splitting  between $N_1$ and $E_1$ in our scans, $m_{E_1} > 1.2 \, m_{N_1}$, the decays of $E_1$ are always prompt. 

\subsection{Constraints on the Charged Higgs}

\noindent
Collider limits on the charged Higgs are mostly relevant for very light masses, namely $m_{H^{\pm}} < m_t$. In this case. light charged Higgs can be searched in the decays, $t \rightarrow H^{\pm} b$ followed by $H^{\pm} \rightarrow c s$ or $H^{\pm} \rightarrow \tau \nu_\tau$. Searches for this processes have been performed both by ATLAS~\cite{Aad:2013hla} and CMS~\cite{Chatrchyan:2012vca,Khachatryan:2015uua}. No sensitive variations in the top branching fractions with respect to the SM have been detected, disfavoring masses of the charged Higgs below 160 GeV. The ATLAS collaboration has performed searches for $H^{\pm}\rightarrow \tau \nu_\tau$~\cite{Aad:2014kga} also in the high mass regime, i.e. $m_{H^{\pm}} > m_t$, with the charged Higgs being produced in association with a top quark, i.e. through the process $g b \rightarrow t H^{\pm}$. The limits obtained, however, cannot yet constrain efficiently most of the 2HDM setups considered in this work (with the possible exception of the Lepton Specific 2HDM), since the $\tau \nu_\tau$ final state has a low branching fraction at high masses~\cite{Akeroyd:2016ymd}. 

\noindent
The mass of the charged Higgs can be also strongly constrained by low energy observables. As these bounds are determined by the value of $\tan\beta$, they are actually dependent on the type of 2HDM realization. For an extensive review we refer, for example, to Ref.~\cite{Akeroyd:2016ymd}. We will instead summarize, in the following, the constraints relevant to our analysis.

\noindent
We have first of all to consider loop induced contributions to the $B\rightarrow X_s \gamma$ process. These depend on the coupling of the charged Higgs to $t$,$b$ and $s$ quarks. In the type-I and lepton specific models, all the relevant couplings are suppressed by $1/\tan\beta$ and, hence, sizable constraints are obtained only for very low $\tan\beta$~\cite{Hermann:2012fc}. Much stronger bounds are instead obtained in 2HDM-II, excluding masses of the charged Higgs up to order of 400 GeV~\cite{Abbott:1979dt,Ciuchini:1997xe,Borzumati:1998tg,Borzumati:1998nx}, practically independent from the value of $\tan\beta$. A second relevant bound comes from the semileptonic decays of the pseudoscalar mesons, in particular $B(B \rightarrow \tau \nu)$. By requiring the ratio $r=B(B \rightarrow \tau \nu_\tau)/B(B \rightarrow \tau \nu_\tau)_{\rm SM}$ to be consistent with the experimental determination $r=1.56 \pm 0.47$~\cite{Agashe:2014kda,Charles:2004jd}, one obtains, only for the type-II 2HDM, a limit on the bidimensional plane $(m_{H^{\pm}},\tan\beta)$ which is relevant for $\tan\beta \gtrsim 20$.

\subsection{Summary of Results}

\begin{figure}[htb]
\begin{center}
\includegraphics[width=6.5 cm]{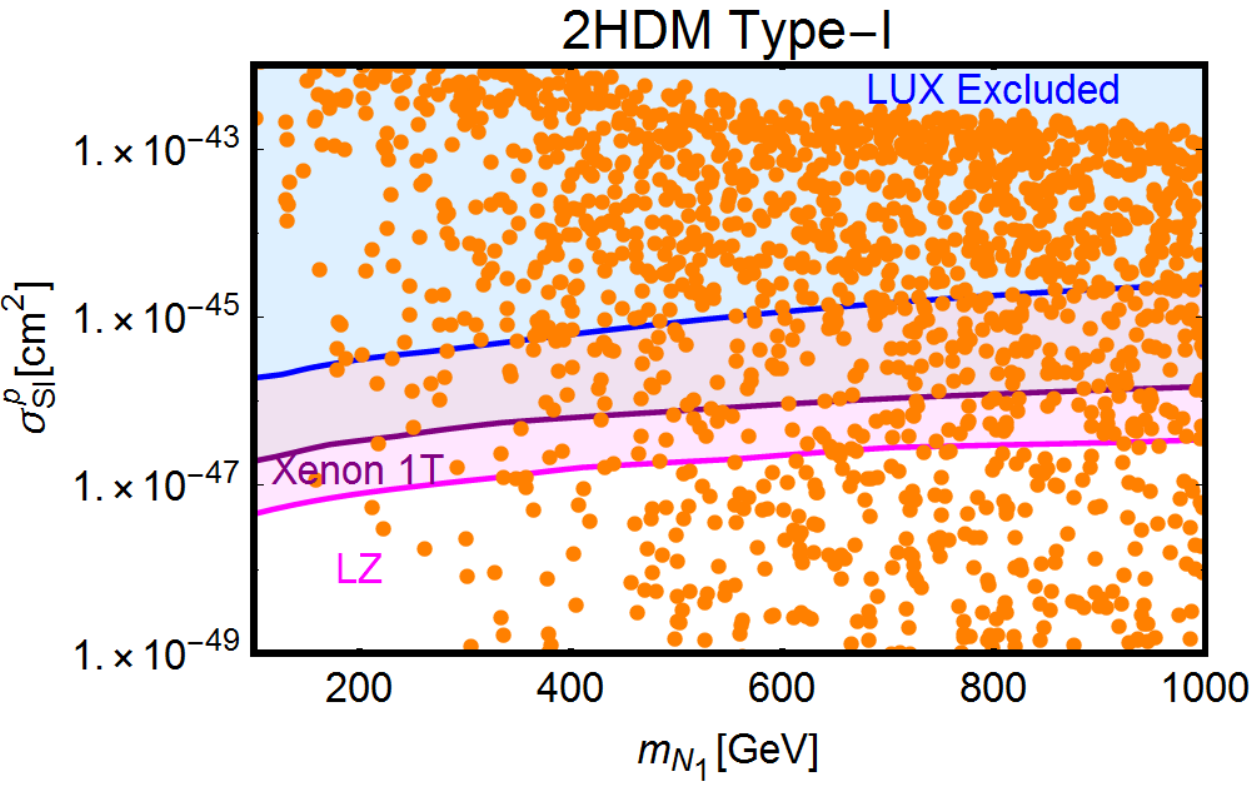}
\includegraphics[width=6.5 cm]{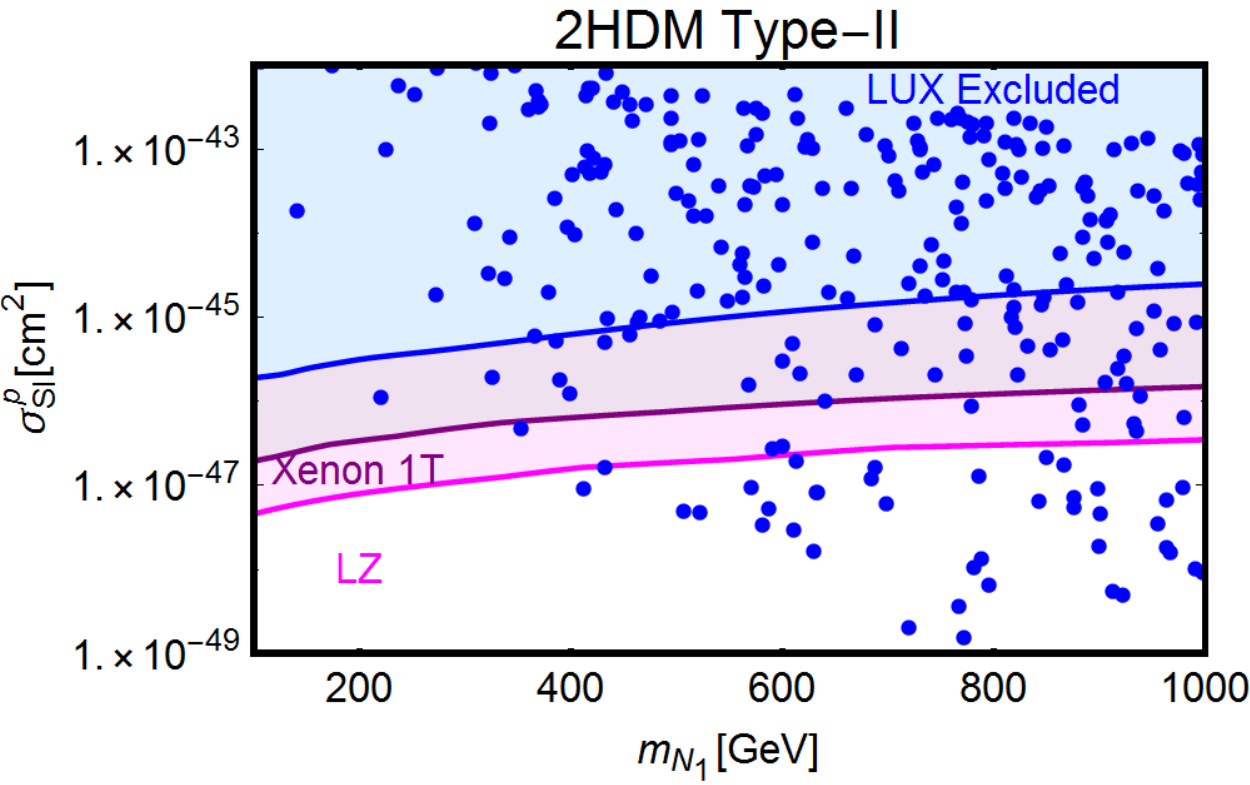}\\
\includegraphics[width=6.5 cm]{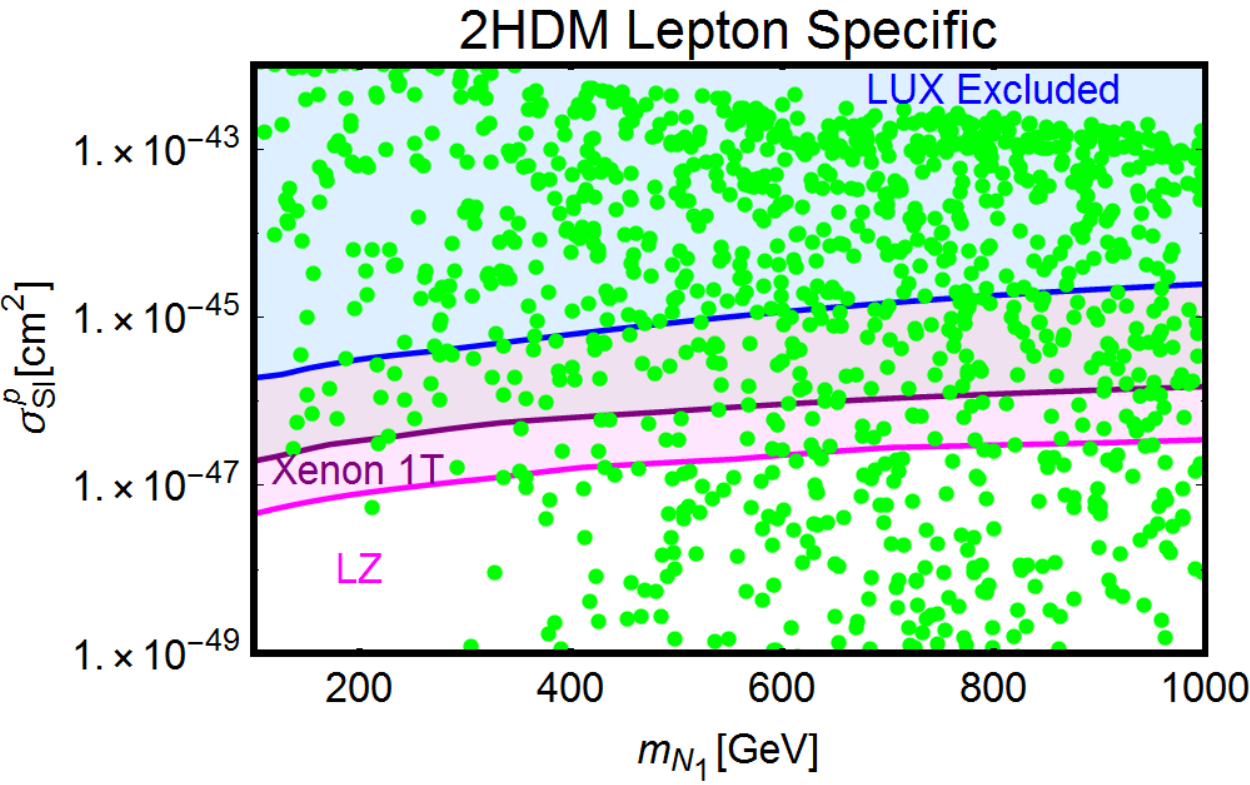}
\includegraphics[width=6.5 cm]{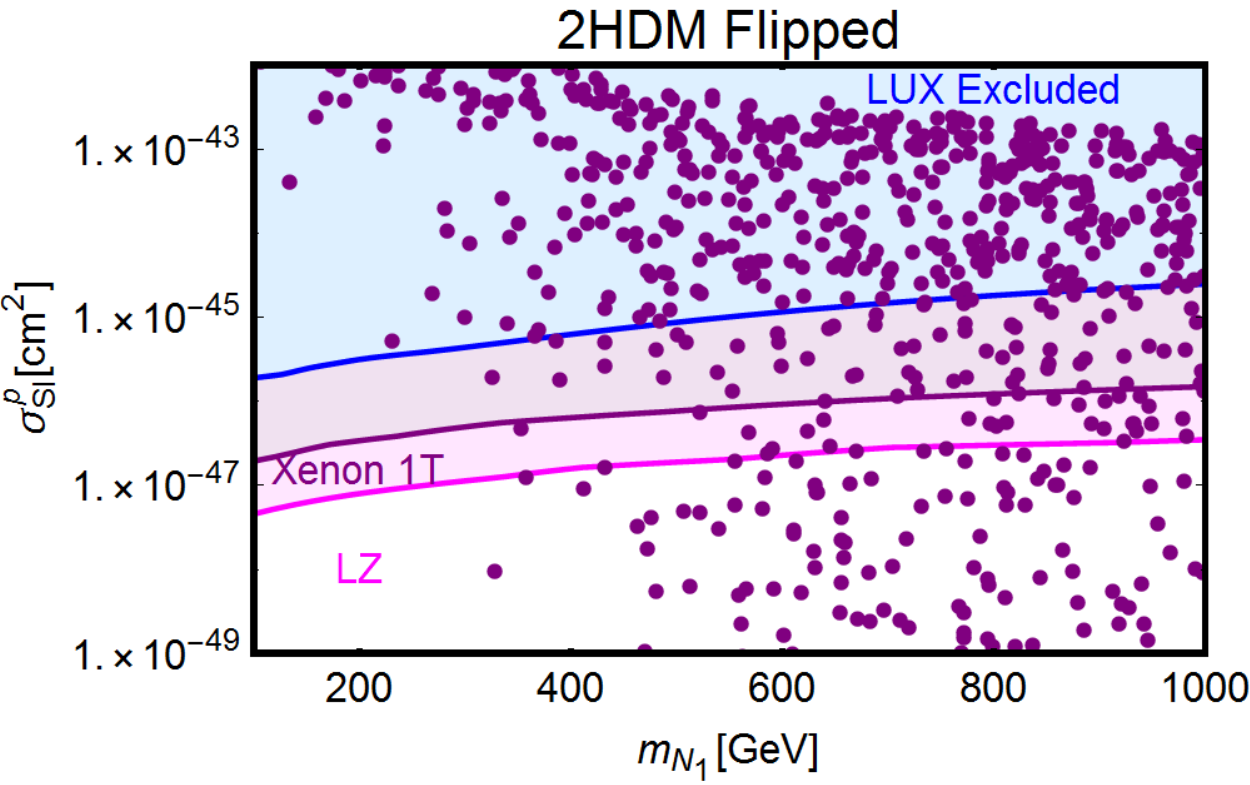}
\end{center}
\caption{Summary plots including all the constraints discussed along this work. Each of the four panels of the figure refers to a different 2HDM realization, indicated on the top of the panel themselves.}
\label{fig:summary}
\end{figure}

\noindent
The results of our study are summarized in fig.~(\ref{fig:summary}). Here we have put all together the results for DM phenomenology with theoretical constraints, i.e. scalar quartic couplings RGEs, EWPT constraints, limits from collider searches, mostly $H/A \rightarrow \tau \tau$, and constraints from low energy observables (for the latter we have adopted the limits on $(m_{H^{\pm}},\tan\beta)$ as reported in Refs.~\cite{Flacher:2008zq,Hermann:2012fc}).

\noindent
The four panels of fig.~(\ref{fig:summary}) show, for each 2HDM realization, namely type I (orange points), type II (blue points), Lepton Specific (green points), Flipped (purple points), in the plane $(m_{N_1},\sigma_{N_1 p}^{\rm SI})$, the model points providing the correct DM relic density and satisfying the constraints listed above.

\noindent
As evident, the main impact of the 2HDM realizations relies on the minimal viable value of the DM mass. Indeed, in the region allowed by the LUX limits, with the exception of (fine tuned, especially for high $\tan\beta$) resonant DM annihilation, the correct relic density is mostly achieved through annihilations involving charged Higgses as final states. Very light charged Higgses are however forbidden, because of constraints from low-energy observables. In particular, type-II 2HDM appears disfavored because of the additional effect of limits from $A/H \rightarrow \tau \tau$ (cfr. fig.~\ref{fig:tautauI}). On the contrary, for type-I and Lepton specific realizations, viable DM  candidates as light as approximately 150 GeV can be achieved.

%%%%%%%%%%%%%%%%%%%%%%%%%%%%%%%%%%%%%%%%%%%%%%%%%%%%%%%%%%%%%%%%%%%%%%%%%%
\section{Conclusions}
%%%%%%%%%%%%%%%%%%%%%%%%%%%%%%%%%%%%%%%%%%%%%%%%%%%%%%%%%%%%%%%%%%%%%%%%%%

\noindent
In this work, we have performed an extensive study of the impact of the addition of a family of vector like fermions, with suitable quantum numbers such as to provide a DM candidate, to the SM and to various types of 2HDMs.

\noindent
The SM+VLLs realization is strongly constrained. The correct relic density implies too strong interactions with the $Z$-boson, ruled out-by DM Direct Detection unless the DM, and hence in turn the whole spectrum of the new fermions, lie above the TeV scale.

\noindent
Lower DM masses can instead be achieved in 2HDM realizations. Indeed, s-channel enhancement, in correspondence with the $H/A$ poles, can provide the correct relic density even for a small hypercharge/SU(2) component of the Dark Matter. In addition, efficient DM annihilations can be also achieved, in the $H^{\pm} H^{\mp}$ and $W^{\pm} H^{\mp}$ final states. The corresponding cross-section is not directly correlated with the DM DD cross-section, such that it would be possible to evade current and even next future bounds. On the other hand the DM relic density depends on the masses of the new Higgs states. Complementary constraints thus come from their experimental searches. Given their dependence on $\tan\beta$ the allowed parameter space actually depends on the type of couplings of the Higgs doublets with the SM fermions.

\noindent
Type-II, and to lesser extent, flipped 2HDMs, are the most constrained since low values of $m_{H^{\pm}}$ (and in turn DM masses) are excluded by low energy observables. Moreover a large part of the type-II parameter space is excluded by limits from searches of $A/H \rightarrow \tau \tau$. Combining these constraints, DM masses below 400 GeV are strongly disfavored. For the other two 2HDM realizations, constraints from searches of extra Higgses are not yet competitive with DM constraints and lower DM masses are accessible.

\noindent
Although the size of the Yukawa couplings of the charged VLLs can account for the correct DM relic density, it does not account for a significant enhancement of the diphoton production rates observable at colliders. This happens because the limit from EWPT and RGE forbid values greater than $\sim 1$ for this couplings.

\noindent
Moreover, the possibility of a direct observation of the VLLs appears similarly contrived.

%%%%%%%%%%%%%%%%%%%%%%%%%%%%%%%%%%%%%%%%%%%%%%%%%%%%%%%%%%%%%%%%%%%%%%%%%%
\section*{Acknowledgements}
%%%%%%%%%%%%%%%%%%%%%%%%%%%%%%%%%%%%%%%%%%%%%%%%%%%%%%%%%%%%%%%%%%%%%%%%%%

\noindent
This project has received partial support from the ERC advanced grant Higgs@LHC. The authors are indebted to Abdelhak Djouadi for inspiration and for careful reading of the draft. They also thank Damir Be\v{c}irevi\'{c} and Gr\'{e}gory Moreau for fruitful discussions, and Nicolas Bizot for collaboration at the early stages of the project. The authors warmly thank the CERN Theory Division for hospitality during part of the completion of this work. 

\bibliographystyle{utphys}
\bibliography{bibfile}{}
%%%%%%%%%%%%%%%%%%%%%%%%%%%%%%

\end{document}